\documentclass[%
 reprint,
 superscriptaddress,
 amsmath,amssymb,
 aps,
 prb,
 floatfix,
]{revtex4-2}

\usepackage[none]{hyphenat}
\usepackage{graphicx}
\usepackage{dcolumn}
\usepackage{bm}
\usepackage[utf8]{inputenc}
\usepackage[T1]{fontenc}
\usepackage{mathptmx}
\usepackage{textcomp}
\usepackage{siunitx}
\usepackage{bigints}
\usepackage{amsmath}
\usepackage{wrapfig}
\usepackage{ragged2e}

\usepackage{hyperref}
\usepackage[switch]{lineno}
\usepackage[table,xcdraw,dvipsnames]{xcolor}
\usepackage{subcaption}
\usepackage{booktabs}
\usepackage{float}
\usepackage{color}
\usepackage{ulem}
\usepackage{multirow}

\begin{document}

\title{Wavenumber-dependent magnetic losses in YIG-GGG heterostructures at millikelvin temperatures}

\author{David Schmoll\href{https://orcid.org/0000-0001-5260-2052}{\includegraphics[scale=0.02]{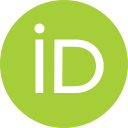}}}
\email{david.schmoll@univie.ac.at}
\affiliation{Faculty of Physics, University of Vienna, Boltzmanngasse 5, 1090, Vienna, Austria}
\affiliation{Vienna Doctoral School in Physics, University of Vienna, Boltzmanngasse 5, 1090, Vienna, Austria}
\author{Andrey A. Voronov\href{https://orcid.org/0000-0002-3887-346X}{\includegraphics[scale=0.02]{orcid.png}}}%
\affiliation{Faculty of Physics, University of Vienna, Boltzmanngasse 5, 1090, Vienna, Austria}
\affiliation{Vienna Doctoral School in Physics, University of Vienna, Boltzmanngasse 5, 1090, Vienna, Austria}
\author{Rostyslav O. Serha\href{https://orcid.org/0009-0007-2962-1109}{\includegraphics[scale=0.02]{orcid.png}}}%
\affiliation{Faculty of Physics, University of Vienna, Boltzmanngasse 5, 1090, Vienna, Austria}
\affiliation{Vienna Doctoral School in Physics, University of Vienna, Boltzmanngasse 5, 1090, Vienna, Austria}
\author{Denys Slobodianiuk}%
\affiliation{Institute of Magnetism, 03142, Kyiv, Ukraine}
\author{Khrystyna~O.~Levchenko\href{https://orcid.org/0000-0002-0903-5942}{\includegraphics[scale=0.02]{orcid.png}}}
\affiliation{Faculty of Physics, University of Vienna, Boltzmanngasse 5, 1090, Vienna, Austria}
\author{Claas Abert\href{https://orcid.org/0000-0002-4999-0311}{\includegraphics[scale=0.02]{orcid.png}}}%
\affiliation{Faculty of Physics, University of Vienna, Boltzmanngasse 5, 1090, Vienna, Austria}
\affiliation{Platform MMM Mathematics–Magnetism–Materials, University of Vienna, 1090}
\author{Sebastian Knauer\href{https://orcid.org/0000-0002-5790-4575}{\includegraphics[scale=0.02]{orcid.png}}}%
\affiliation{Faculty of Physics, University of Vienna, Boltzmanngasse 5, 1090, Vienna, Austria}
\author{Dieter Suess\href{https://orcid.org/0000-0001-5453-9974}{\includegraphics[scale=0.02]{orcid.png}}}%
\affiliation{Faculty of Physics, University of Vienna, Boltzmanngasse 5, 1090, Vienna, Austria}
\affiliation{Platform MMM Mathematics–Magnetism–Materials, University of Vienna, 1090}
\author{Roman Verba\href{https://orcid.org/0000-0001-8811-6232}{\includegraphics[scale=0.02]{orcid.png}}}%
\affiliation{Institute of Magnetism, NAS of Ukraine and MES of Ukraine, 03142, Kyiv, Ukraine}
\author{Andrii V. Chumak\href{https://orcid.org/0000-0001-5515-0848}{\includegraphics[scale=0.02]{orcid.png}}}%
\email{andrii.chumak@univie.ac.at}
\affiliation{Faculty of Physics, University of Vienna, Boltzmanngasse 5, 1090, Vienna, Austria}
\date{\today}

\begin{abstract}

Magnons have inspired potential applications in modern quantum technologies and hybrid quantum systems due to their intrinsic nonlinearity, nanoscale scalability, and a unique set of experimentally accessible parameters for manipulating their dispersion. Such magnon-based quantum technologies demand long decoherence times, millikelvin temperatures, and minimal dissipation. Due to its low magnetic damping, the ferrimagnet yttrium iron garnet (YIG), grown on gadolinium gallium garnet (GGG), is the most promising material for this objective. To comprehend the magnetic losses of propagating magnons in such YIG-GGG heterostructures at cryogenic temperatures, we investigate magnon transport in a micrometer-thick YIG sample via propagating spin-wave spectroscopy (PSWS) measurements for temperatures between $\SI{4}{\kelvin}$ to $\SI{26}{\milli\kelvin}$. We demonstrate an increase in the dissipation rate with wavenumber at cryogenic temperatures, caused by dipolar coupling to the partially magnetized GGG substrate. Additionally, we observe a temperature-dependent decrease in spin-wave transmission, attributed to rare earth ion relaxations. The critical role of the additional dissipation channels at cryogenic temperatures is underpinned by the comparison of the experimental results with theoretical calculations and micromagnetic simulations. Our findings strengthen the understanding of magnon losses at millikelvin temperatures, which is essential for the future detection of individual propagating magnons.

\end{abstract}

\maketitle

\section{Introduction}
\label{sec:introduction}

The research field of magnonics deals with spin waves and their quanta called magnons, as the eigenexcitations of the collective spins of magnetically ordered media and investigates their application for efficient data transmission and processing. Between 1960 and 1980 a wide range of devices for analog signal processing were developed, mainly based on the insulating ferrimagnet yttrium iron garnet (YIG), due to the material's uniquely low magnetic damping~\cite{Serga2010, Adam1988, Glass1988, Ishak1988, Morgenthaler1988, Rodrigue1988}. Continuous advances in nanotechnology and the demand for innovation in data transportation and processing sparked interest in the potential of spin waves as data carriers in novel computing schemes~\cite{Zeenba2024, Finocchio2024, Wang2024, Wang2023, Pirro2021, Mahmoud2020, Heinz2020} and their utilization in wide frequency bandwidth RF devices~\cite{WangNature, Heussner2020, Vogt2014, Yttrium2009}.

Attention from inside and outside the magnonics community also focuses on the quantum nature of magnons, due to the intrinsic nonlinearity and the scalability down to the nanoscale~\cite{Chumak2022, Barman2021}. The diversity of spin-wave dispersion characteristics and the wide range of experimentally accessible parameters, such as bias field, sample shape, or spin-wave frequency, which allow the manipulation of the dispersion, offer exciting opportunities for the coupling of magnons to phonons, superconducting qubits, and photons~\cite{Zhang2023, Jiang2023, Baity2021, Li2020, Quirion2019}. Such hybrid quantum systems combine physical platforms that are well suited to perform different tasks in applications for quantum information, communication, and sensing~\cite{Forsch2020, Kurizki2015}.

To maintain the coherence of the quantum states, strong coupling between the subsystems and a low decoherence rate are essential. Due to the low dissipation and the large spin density, allowing strong coupling to other quantum systems~\cite{Cao2015}, YIG is the material of choice for the investigation of magnonic quantum states. First steps have already been achieved, by the coupling of a superconducting qubit and a single uniform magnon in a bulk YIG sphere~\cite{Nakamura2020, Tabuchi2015}. However, to fully utilize spin waves at the quantum level, the transition from uniform precessions to propagating single magnons with spacially separated sources and detectors, poses the next big challenge.

Classical propagating spin waves in a YIG film have been shown at temperatures down to \SI{45}{\milli\kelvin}~\cite{Knauer2023}. Such ultralow temperatures are necessary to exclude the presence of thermally excited magnons. However, the majority of YIG films are grown on the substrate gadolinium gallium garnet (GGG) to ensure high-quality crystals due to the close lattice matching. Ferromagnetic resonance~(FMR) studies reported, that GGG is exhibiting paramagnetic behaviour, acts as a damping source for magnons~\cite{Kosen2019, Danilov1989, Guo2022}, and, in the case of in-plane magnetized films, generates an inhomogeneous magnetic stray field antiparallel to the applied external field, at decreasing temperatures~\cite{Serha2024}. The understanding of the behavior of the dissipation rate in the YIG/GGG magnetic system, not only for FMR but also for propagating magnons, is important for investigations of magnonic quantum states and their coupling to other quantum systems.
\begin{figure*}[tb]
	\includegraphics[width=\textwidth]{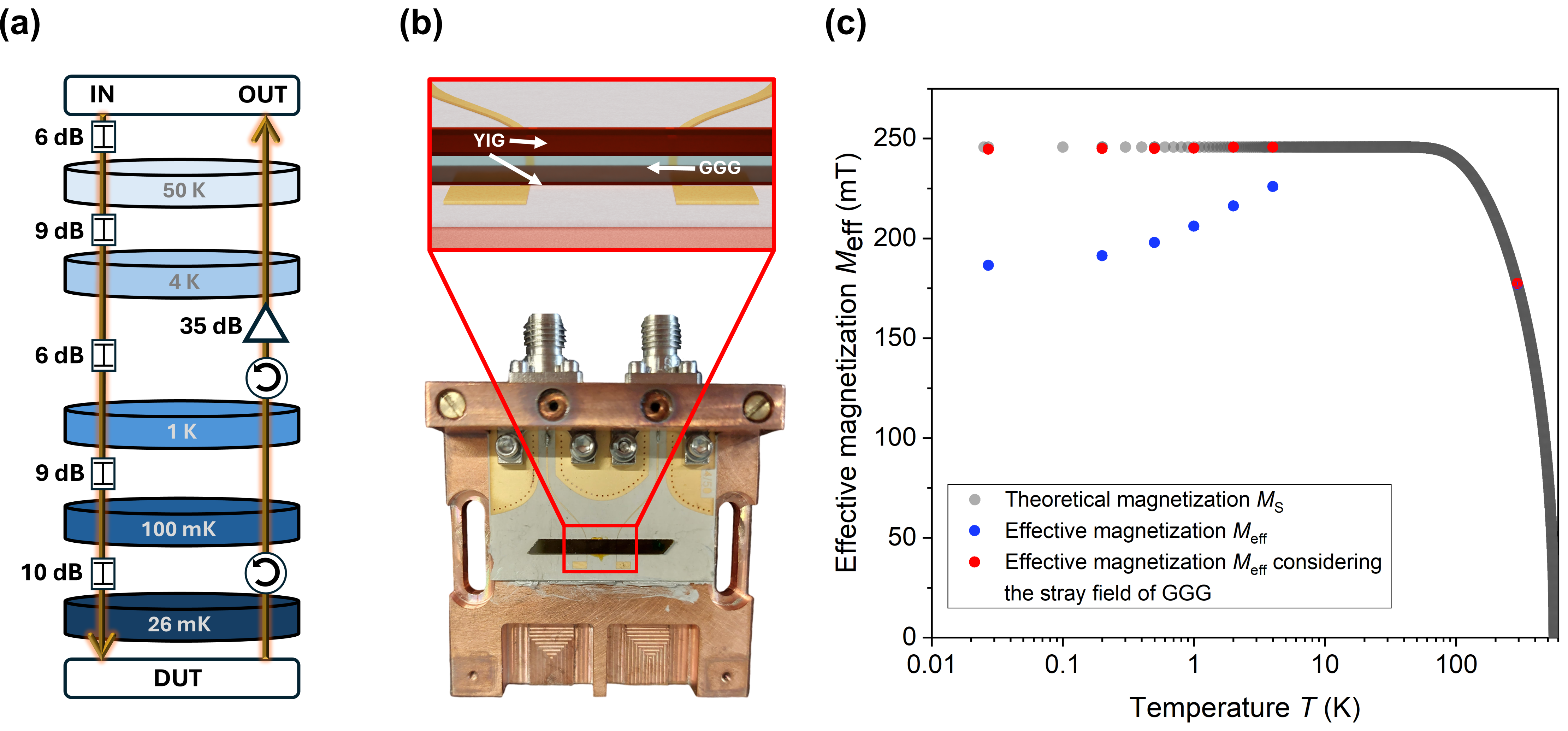}
    \captionsetup{justification=justified}
	\caption{\justifying (a) Schematic diagram of the dilution refrigerator transmission line assembly. The input and the output line are connected to a vector network analyzer (VNA). (b) Illustration of the investigated YIG/GGG sample, placed on top of two microstrip antennas, which are separated by \SI{4}{\milli\meter}. The GGG substrate is illustrated in transparent light blue, with a dark brown YIG film on both sides. (c) Theoretical saturation magnetization of YIG (grey) and effective magnetization extracted from the experimental FMR frequency and the external magnetic fields applied at the different temperatures, without (blue) and with consideration of the magnetic stray field of GGG (red). The magnetization error lies within the size of the dots.} 
	\label{fig:fig01}
\end{figure*}

Here, we report on propagating spin-wave spectroscopy~(PSWS) measurements at temperatures between \SI{4}{\kelvin} and \SI{26}{\milli\kelvin} in a micrometer-thick YIG sample, grown on a GGG substrate. We developed a method to extract the wavenumber dependent dissipation rate $\Gamma_\mathrm{k}$ from the experimental spin-wave transmission spectra and the semi-analytically calculated dispersion relation of magnetostatic waves in a dipolar coupled layered magnetic structure. Additionally, the measurements and the theoretical calculations are compared to micromagnetic simulations. Our experiments reveal a significant change of the dissipation with wavenumber at decreasing temperatures, with an increase of $\Gamma_\mathrm{k}$ up to $55\%$ for the excited wavenumbers between $0$ and $450$~rad/cm at $\SI{26}{\milli\kelvin}$. In agreement with literature~\cite{Weides2018, Mihalceanu2018, Seiden1964, Sparks1961, Dillon1959, Spencer1959}, we also observe increased magnetic losses due to rare-earth ion relaxations at \SI{4}{\kelvin}. As this dissipation channel is suppressed at sub-Kelvin temperatures~\cite{Kosen2019}, $\Gamma_\mathrm{k}$ decreases by further lowering the temperature. Below \SI{500}{\milli\kelvin} the magnetic losses do not change anymore, due to the frustrated spin-system of GGG, resulting in a complex phase transition at such low temperatures~\cite{Serha2024, Petrenko1998, Tsui1999, Schiffer1994, Deen2015}. 

\section{Methods}
\label{sec:modeling}
\subsection{\label{subsec:micromag}Experimental methods}

We used a LPE-grown <111>~-~orientated YIG film of $\SI{7.78}{\micro\meter}$ thickness, grown on a $\SI{500}{\micro\meter}$-thick GGG substrate. The sample is $\SI{20}{\milli\meter}$ long, $\SI{2}{\milli\meter}$ wide, and has two $30^\circ$ edge cuts to minimize reflections. An antenna printed circuit board (PCB) with two $\SI{50}{\micro\meter}$ wide microstrips, separated by $\SI{4}{\milli\meter}$, was used for PSWS measurements at room temperature and at cryogenic temperatures (see Fig.~\ref{fig:fig01}(b)). At room temperature the sample was placed between a GMW variable pole gap electromagnet and connected to a $\SI{40}{\giga\hertz}$ Rohde \& Schwarz ZVA 40 VNA. The cryogenic setup is based on a Bluefors-LD250 dilution refrigerator, capable to reach base temperatures as low as $\SI{10}{\milli\kelvin}$, and a shored AMI superconducting vector magnet. A $\SI{70}{\giga\hertz}$ Anritsu MS4647B VNA is connected to the antennas via a $\SI{40}{\decibel}$ attenuated input line and a superconducting output line, equipped with a RF circulator at the mixing chamber and at $\SI{4}{\kelvin}$. The attenuation at the input line and the RF circulators at the output line are necessary to isolate the sample from thermal noise. To increase the signal to noise ratio (SNR), a LNF LNC 2-6 A RF amplifier with $\SI{35}{\decibel}$ gain at the frequency of $\SI{4.5}{\giga\hertz}$ is placed at the $\SI{4}{\kelvin}$ stage in the dilution refrigerator (see Fig.~\ref{fig:fig01}(a)).

Spin waves were measured in the Magnetostatic Surface Spin Wave (MSSW) configuration, by positioning the sample in the homogeneous region of the applied magnetic field in both setups and magnetizing the sample in plane and perpendicular to the propagation direction of the spin wave. Magnons are investigated at the YIG surface, which is in touch with air, with an input power of $\SI{-40}{\decibel}$ at the antenna. Spin-wave propagation is detected as the $S_{21}$ transmission spectrum via the VNA. Measurements were performed at $\SI{293}{\kelvin}$ and in the temperature range between $\SI{4}{\kelvin} - \SI{26}{\milli\kelvin}$. At $\SI{26}{\milli\kelvin}$ the dilution refrigerator has a cooling power of $\SI{14}{\micro\W}$ and is still able to keep the system and the inserted sample in thermal equilibrium. To ensure the measurements have the same signal floor, transmission losses up to the antenna PCB were measured in both setups and subtracted from the transmission spectra.

Figure~\ref{fig:fig01}(c) depicts the effective magnetization, already accounting for demagnetization as well as anisotropy fields and calculated with the Kittel-formula for the FMR-frequency of $\SI{4.515}{\giga\hertz}$, as the overlapping point of the experimentally obtained transmission spectra. By considering the alteration of the applied external field due to the stray field of GGG, the effective magnetization is in agreement with the theoretically predicted saturation magnetization of YIG. The temperature dependent magnitude of the GGG stray field can be calculated by following the procedure presented in \cite{Serha2024, Knauer2023}.

To compare the spin-wave transmission curves at different temperatures, we overlap the spectra at the FMR point in the frequency domain. Hence, the applied external field at the different temperatures was adjusted, to account for the frequency shift introduced by the increasing saturation magnetization of YIG~\cite{Cherepanov1993} and the antiparallel stray field, created by the partially magnetized GGG substrate~\cite{Serha2024}. 

\subsection{Semi-analytical dispersion calculations}
\label{sec:modelingSL}

The PSWS measurements in the room temperature and dilution refrigerator setup allow to experimentally record spin-wave transmission in the frequency domain at different temperatures. To investigate the dependency of magnon transport on wavenumber $k$ at different temperatures, the experimentally measured transmission $S_{21}$ needs to be converted from the frequency domain to the wavenumber domain via the dispersion relation. Here, we adopt the method of magnetostatic wave dispersion calculation in multilayered structures, developed in \cite{Emtage1984}, to our system. As we are interested not only in the dispersion but also in the damping rate of magnetostatic waves, we introduce the complex wave frequency $\tilde\omega_\mathrm{k}~=~\omega_\mathrm{k}~+~i \Gamma_\mathrm{k}$, where the real part ($\omega_\mathrm{k}$) describes the dispersion relation and the imaginary part ($\Gamma_\mathrm{k}$) denotes the damping rate. 

YIG and GGG layers are described by their magnetic permeability tensors with hermitian and anti-hermitian components
\begin{equation}
    \mu_\mathrm{n} = \frac{\tilde\omega_\mathrm{k}^2 - \omega_\mathrm{\bot,n}^2}{\tilde\omega_\mathrm{k}^2 - \omega^2_\mathrm{H, n}} \,, \quad 
    \mu_\mathrm{a, n} = \frac{\tilde\omega_\mathrm{k} \omega_\mathrm{M, n}}{\tilde\omega_\mathrm{k}^2 - \omega^2_\mathrm{H, n}} \,,
\end{equation}
where the index $n$ represents the YIG or GGG layer. The quantities in these equations are introduced as:
\begin{subequations}\label{e:wH}
    \begin{equation}
        \omega_\mathrm{H, YIG} = \gamma\mu_0 (H_0 + H_\mathrm{a}) + i\alpha_\mathrm{G} \omega_\mathrm{k} \,,
    \end{equation}
    \begin{equation}
        \omega_\mathrm{H, GGG} = \gamma\mu_0 H_0 + i\gamma\mu_0 \frac{\Delta H}{2} \,,
    \end{equation}
\end{subequations}
\begin{equation}
    \omega_\mathrm{M, n} = \gamma\mu_0 M_\mathrm{n} \,, \quad
    \omega_\mathrm{\bot, n} = \sqrt{\omega_\mathrm{H, n} (\omega_\mathrm{H, n} + \omega_\mathrm{M, n})} \,,
\end{equation}
with $M_\mathrm{YIG}$ being the saturation magnetization for YIG, $M_\mathrm{GGG}$ as the net static magnetization of GGG at a given field and temperature, and $H_\mathrm{a}$ being the weak effective anisotropy field in YIG (see below). 

Equations~\ref{e:wH}(a) and \ref{e:wH}(b) imply that we are using different dissipation models for YIG and GGG. For YIG, we use the standard Gilbert model with $\alpha_\mathrm{G}$ being the Gilbert damping parameter. Its room-temperature value was determined experimentally using a standard FMR setup, with the instrumentation of the room-temperature PSWS setup and a commercial coplanar waveguide (CPW) FMR antenna from NanOsc Instruments. The recorded FMR spectra revealed a Gilbert damping of $\alpha_{293\mathrm{K}}~=~(5.2~\pm~2.5)\cdot 10^{-5}$. 

The nature of FMR linewidth in GGG is different and shows inhomogeneous linewidth broadening due to atomic-scale nonuniform stochastic effective fields of rare-earth ions. This broadening is weakly dependent on the temperature and GGG magnetization and is often accounted for by a single parameter  $\mu_0\Delta H \approx $ \SI{400}{\milli\tesla}~\cite{Barak1992}. Although this approach is simplified and phenomenological, it provides good qualitative agreement with the experimental data as will be shown below. 

Boundary conditions for magnetic field and induction in the air-YIG-GGG-air structure result in the following secular equations, describing the spin wave spectrum in the MSSW geometry:
\begin{equation}
     D_\mathrm{YIG} D_\mathrm{GGG} + E_\mathrm{YIG} E_\mathrm{GGG} = 0.
\label{Dispersion}
\end{equation}
where:
\begin{subequations}\label{e:def-coef}
    \begin{equation}
         D_\mathrm{YIG} = 2\mu_\mathrm{YIG} + (1 + \mu^2_\mathrm{YIG} - \mu^2_\mathrm{a, YIG})\tanh{(k d_\mathrm{YIG})} \,,
    \end{equation}
    \begin{equation}
         D_\mathrm{GGG} = 2\mu_\mathrm{GGG} + (1 + \mu^2_\mathrm{GGG} - \mu^2_\mathrm{a, GGG})\tanh{(k d_\mathrm{GGG})} \,,
    \end{equation}
    \begin{equation}
         F_\mathrm{YIG} = (-1 + \mu^2_\mathrm{YIG} - \mu^2_\mathrm{a, YIG} - 2\mu_\mathrm{a, YIG})\tanh{(k d_\mathrm{YIG})} \,,
    \end{equation}
    \begin{equation}
         E_\mathrm{GGG} = (1 - \mu^2_\mathrm{GGG} + \mu^2_\mathrm{a, GGG} - 2\mu_\mathrm{a, GGG})\tanh{(k d_\mathrm{GGG})} \ .
    \end{equation}
\end{subequations}

Here, $d_\mathrm{n}$ is the thickness of the layers: $d_{\mathrm{YIG}}~=~\SI{7.78}{\micro\meter}$ and $d_{\mathrm{GGG}}~=~\SI{500}{\micro\meter}$. Equation~\ref{Dispersion} is solved numerically to obtain the values of the magnetostatic wave frequency $\omega_\mathrm{k}$ and dissipation rate $\Gamma_\mathrm{k}$ at a given wavenumber $k$. Note, that Eqs.~\ref{e:def-coef}(a)-(d) are valid for $k~>~0$ only~\cite{Emtage1984}. For the investigation of wave dispersion in the opposite direction, one should virtually reverse the bias magnetic field, which results in the sign reversal of the anti-hermitian components $\mu_{\mathrm{a,n}}$ of the magnetic permeability tensor for both the YIG and GGG layers.

\subsection{Numerical simulations}
\label{sec:numsim}

To support the quasi-analytical results, the dispersion relation was computed using micromagnetic simulations. The experimental configuration consisted of a ferrimagnetic YIG layer with a thickness of \(\SI{7.78}{\micro\meter}\) and a paramagnetic GGG substrate with a thickness of \(\SI{500}{\micro\meter}\). The two orders of magnitude difference in thickness makes simulating the entire GGG region in one micromagnetic model infeasible, while maintaining uniform discretization along the vertical axis.

\begin{figure}
    \centering
    \includegraphics[width=0.98\linewidth]{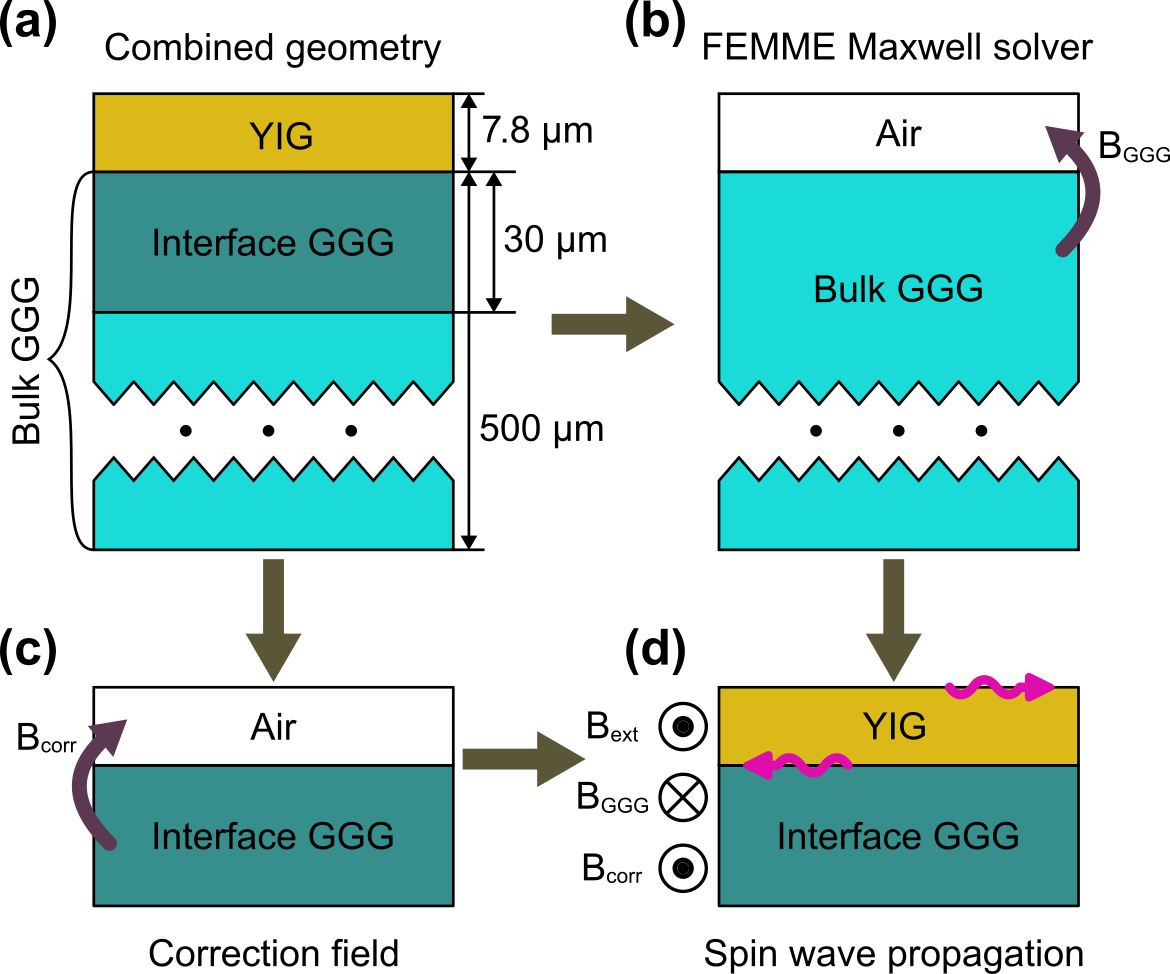}
    \captionsetup{justification=justified}
    \caption{\justifying Schematic representation of the numerical simulation approach for the computation of the spin-wave dispersion and lifetime. (a)~Splitting of the simulation into multiple domains, including the introduction of the “interface GGG” region, to account for the dynamic influence of the partially magnetized GGG on YIG. (b)~Calculation of the static stray field $B_\mathrm{GGG}$ induced by the bulk \(\SI{500}{\micro\meter}\)-thick GGG substrate, using the nonlinear Maxwell solver in FEMME. (c)~Simulation of the correction field $B_\mathrm{corr}$, resulting from the inclusion of an “interface GGG” layer into the micromagnetic simulations. This field is subtracted from the YIG simulation to eliminate parasitic effects. (d)~Final micromagnetic simulation configuration for the dispersion calculation in the layered magnetic structure, incorporating the stray field $B_\mathrm{GGG}$ and the correction field $B_\mathrm{corr}$.}
    \label{fig:numerics}
\end{figure}

To address this challenge, the GGG substrate was split into distinct domains, as shown in Fig.~\ref{fig:numerics}(a), separating its static and dynamic influences on spin-wave propagation in YIG. The static interaction, governed by the stray field generated by the partially magnetized GGG (Fig.~\ref{fig:numerics}(b)), was calculated using the FEMME software, which solves the nonlinear Maxwell equations for the temperature and field-dependent magnetization of GGG. This finite element/boundary element method employed a mesh size of \(\SI{0.1}{\milli\meter}\)~\cite{Dieter2012}. The resulting stray field $B_\mathrm{GGG}$ was evaluated within a \(\SI{7.78}{\micro\meter}\)-thick layer above the GGG and incorporated into the micromagnetic simulations (Fig.~\ref{fig:numerics}(d)) as an additional constant bias field, capturing the static substrate effects on the magnetic phenomena in YIG.

The micromagnetic simulations of the dispersion in the YIG film were performed using the GPU-accelerated software \texttt{magnum.np}~\cite{bruckner2023magnum} (Fig.~\ref{fig:numerics} (d)). The simulated geometry consists of a YIG film of size \(\SI{4}{\milli\meter} \times \SI{2}{\milli\meter} \times \SI{7.78}{\micro\meter}\), discretized into cells of \(\SI{5}{\micro\meter} \times \SI{20}{\micro\meter} \times \SI{200}{\nano\meter}\). To account for dynamic interactions, an additional \(\SI{30}{\micro\meter}\)-thick GGG-interface layer was introduced beneath the YIG. This additional layer allowed coupling between the materials via dipole-dipole interactions at their interface. Moreover, as the GGG’s net magnetization increases with decreasing temperature, spin-wave energy starts to be transferred into the substrate's spin-assembly, altering its dispersion characteristics. However, including the \(\SI{30}{\micro\meter}\)-thick GGG layer introduces an additional static field $B_\mathrm{corr}$, acting on the YIG film, which is redundant as static effects are already accounted for in the Maxwell simulations. To address this, a calibration simulation was performed (Fig.~\ref{fig:numerics}(c)). In this step, only the \(\SI{30}{\micro\meter}\)-thick GGG layer was modeled, and the generated stray field inside a surrounding air box, corresponding to the YIG thickness, was calculated. This field was subtracted as a correction term $B_\mathrm{corr}$, ensuring accurate results in the dispersion simulation (Fig.~\ref{fig:numerics}(d)).

The net magnetization $M_\mathrm{YIG}$ values at specific temperatures and applied fields were taken from Fig.~\ref{fig:fig01}(c), while $\alpha_{293\mathrm{K}}$ was set to the experimentally obtained value of $5.2 \cdot 10^{-5}$. For the \(\SI{30}{\micro\meter}\)-thick substrate layer, the nonuniform broadening of the resonance linewidth in GGG~\cite{Barak1992} was incorporated as the frequency-dependent damping $\alpha_\mathrm{GGG}$, recalculated using the definition of the linewidth in the Gilbert model~\cite{kalarickal2006ferromagnetic}. The exchange constant \(A_\mathrm{YIG}~=~\SI{3.5}{\pico\joule/\meter}\) was used for YIG, while the paramagnetic GGG substrate was assumed to show no exchange interaction ($A_\mathrm{GGG}~=~0$). Using the time-dependent magnetization configuration obtained from micromagnetic simulations, the magnon lifetime $\tau$ was determined by dividing the decay length $\lambda$ for various wavenumbers by the group velocity, as the derivative of Eq.~\eqref{Dispersion}. Numerically, the dispersion was calculated through a $sinc$-pulse excitation at the center of the mesh, followed by FFT analysis. In the magnetostatic surface wave geometry, waves are localized near the YIG surfaces (specifically, the YIG/air surface in the experimental configuration) and as a result, only a thin layer of the partially magnetized GGG substrate near the interface with YIG contributes significantly to the dynamic coupling. Hence, the separation of the GGG layers influence into the spin-wave dispersion into static and dynamic contributions is justified. The developed method incorporates the substrate’s influence while maintaining computational feasibility.

\section{Results and Discussion}
\label{sec:results}
\subsection{Temperature dependent magnon transport}
\label{sec:resultsSKL}
To compare magnon transport at different temperatures and investigate the dependency on the wavenumber $k$, spin-wave transmission was measured at $\SI{293}{\kelvin}$ in the room temperature setup and at $\SI{4}{\kelvin}$, $\SI{2}{\kelvin}$, $\SI{1}{\kelvin}$, $\SI{500}{\milli\kelvin}$, $\SI{200}{\milli\kelvin}$, and $\SI{26}{\milli\kelvin}$ in the dilution refrigerator setup. 
\begin{figure*}[tb]
	\includegraphics[width=\textwidth]{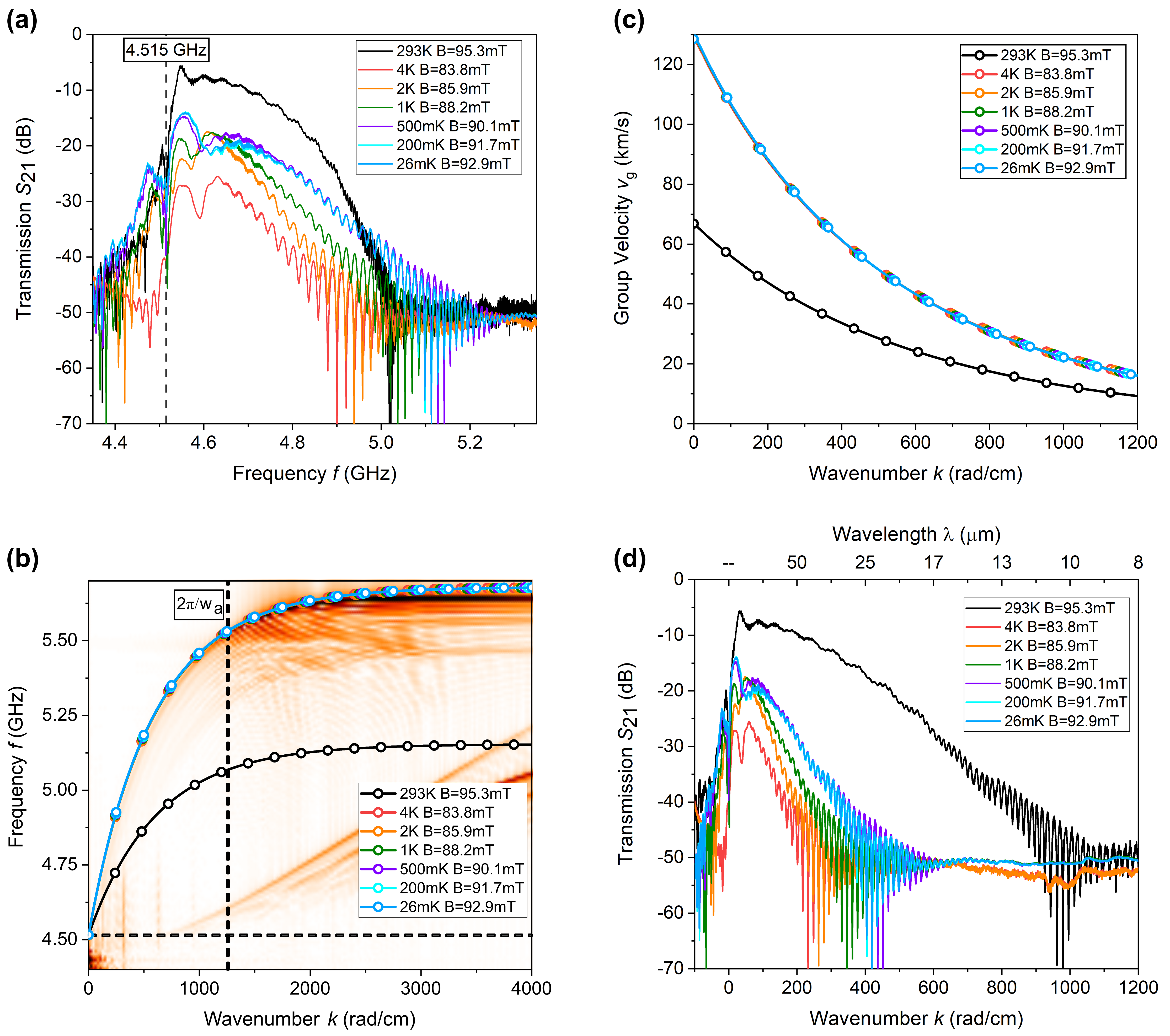}
    \captionsetup{justification=justified}
	\caption{\justifying (a) Spin-wave transmission spectra detected in the MSSW configuration at different temperatures between \SI{293}{\kelvin} and \SI{26}{\milli\kelvin}. To compare magnon transport at different temperatures, the spectra were recorded at different external magnetic fields as indicated in the legend, to overlap them at the FMR frequency. The spectra are corrected for transmission losses down to the microstrip antennas. (b) Dispersion relation calculated for MSSW in a dipolar coupled YIG-GGG bilayer at the experimental temperatures and external magnetic fields (solid lines). Below \SI{4}{\kelvin} the dispersion does not change significantly and the resulting curves overlap. The calculations are compared to micromagnetic simulations, visualized in the color map. The dashed lines highlight the maximal wavenumber, which can be excited with the used antenna. (c) The group velocity $v_\mathrm{g}$, calculated as the derivative of the dispersion relation. (d) Magnon transport in $k$-space, extracted from the experimental spin-wave spectra in the frequency domain via the dispersion relation.} 
	\label{fig:fig02}
\end{figure*}

The recorded $S_{21}$ transmission spectra are depicted in Fig.~\ref{fig:fig02}(a). To analyze the spin-wave signals for the same frequencies, we overlap the curves at the FMR-point of $\SI{4.515}{\giga\hertz}$ (dashed black line) and adjust the external magnetic field accordingly, as noted in the legend of Fig.~\ref{fig:fig02}(a). Compared to the applied field of $\SI{95.3}{\milli\tesla}$ at $\SI{293}{\kelvin}$, the biggest adjustment was necessary at $\SI{4}{\kelvin}$ with an applied field of $\SI{83.8}{\milli\tesla}$, while the smallest difference occurred at $\SI{26}{\milli\kelvin}$ with $\SI{92.9}{\milli\tesla}$. The bigger adjustment of the external field at $\SI{4}{\kelvin}$ is necessary, as the effective magnetization of YIG is already saturated, which increases the spin-wave frequency. However, the magnetic stray field of the GGG substrate at this temperature is still relatively small (see Fig.~\ref{fig:fig01}(c)). By decreasing the temperature further, the magnetization of the GGG substrate increases, which leads to a stronger GGG-induced magnetic stray field~\cite{Serha2024}, while the magnetization of YIG is saturated and does not change anymore with temperature. As the stray field of GGG is antiparallel to the external bias field in the MSSW configuration, it compensates a significant part of the frequency shift due to the higher effective magnetization of YIG and therefore the bias field has to be changed less compared, to the room temperature measurement. We observe the strongest signal for the measured temperatures in the frequency domain at $\SI{293}{\kelvin}$ and the smallest at $\SI{4}{\kelvin}$. By decreasing the temperature further, the spin-wave amplitude increases again until it reaches a plateau at \SI{500}{\milli\kelvin}.

At room temperature, the primary dissipation channels for spin-waves are intrinsic processes, for example, interaction with phonons or two-magnon scattering, which are strongly suppressed in high-quality YIG samples~\cite{Kasuya1961}. Although it can be expected that these intrinsic damping sources decrease with temperature, studies show an increase in the resonance linewidth of rare-earth iron garnets. This behavior is already extensively discussed in literature and is due to the relaxation of rare-earth ions in the crystal lattice~\cite{Weides2018, Mihalceanu2018, Seiden1964, Sparks1961, Dillon1959, Spencer1959}. The increase in linewidth reaches a maximum at approximately $\SI{50}{\kelvin}$ for single crystal YIG spheres and afterwards decreases again with temperature. Note, that the temperature at which the maximum linewidth is observed, strongly depends on the type and concentration of rare-earth ions~\cite{Seiden1964,Dillon1959}. Despite being already below $\SI{50}{\kelvin}$, the stronger decrease in spin-wave amplitude at $\SI{4}{\kelvin}$ can still be mainly attributed to the relaxation of rare-earth ions. As reported in experiments, the linewidth steadily decreases after these temperature peak processes, and the contribution of rare-earth ion relaxation to the magnetic damping is not dominant anymore at sub-Kelvin temperatures~\cite{Spencer1959}. It was shown, that the linewidth decreases even below the room temperature values for substrate-free YIG samples at millikelvin temperatures and input power levels high enough to saturate two-level fluctuation processes~\cite{Kosen2019, Tabuchi2014}. However, in the presence of GGG, the linewidth at millikelvin temperatures is wider than at room temperature, indicating higher damping due to the paramagnetic character of the substrate~\cite{Kosen2019}. Alongside interactions between the YIG and GGG layers close to their interface, we relate this increased damping for uniform precessions partially to the highly inhomogeneous stray field of GGG~\cite{Serha2024}, which leads to smaller resonance frequencies closer to the sample edges, causing asymmetric broadening of the FMR linewidth at decreasing temperatures. Although the GGG related dissipation is most distinct at temperatures below \SI{1}{\kelvin}, it is also already present in the measurements at $\SI{4}{\kelvin}$ (see~\cite{Serha2024, Danilov1989}).

In the frequency domain, the broadest $S_{21}$-spectrum was recorded at $\SI{26}{\milli\kelvin}$, as, compared to room temperature, the dispersion becomes steeper with decreasing the temperature, due to the higher saturation magnetization of YIG (see Fig.~\ref{fig:fig01}(c)). Note, the narrowest spin-wave spectrum was observed at $\SI{4}{\kelvin}$, although the calculated dispersion does not change compared to $\SI{26}{\milli\kelvin}$. We attribute this narrower spectrum width to the significantly increased dissipation at $\SI{4}{\kelvin}$. Figure~\ref{fig:fig02}(b) depicts the dispersion relation calculated with Eq.~\ref{Dispersion} (solid lines) at the different measurement temperatures and considering the altered external field due to the GGG substrate. As the saturation magnetization of YIG is almost constant below $\SI{100}{\kelvin}$ and the stray field and magnetization of GGG is included in the calculation, the dispersion does not change significantly between $\SI{4}{\kelvin}$ and $\SI{26}{\milli\kelvin}$. In order to visualize all the curves in this temperature range, points were added to the theoretically calculated dispersion for clarity. To start at the FMR frequency of $\SI{4.515}{\giga\hertz}$, the quasi-analytical dispersion was fitted to the experimental results with the introduction of a temperature dependent effective anisotropy field $\mu_0 H_\mathrm{a}$, ranging from \SI{1}{\milli\tesla} at \SI{293}{\kelvin} to the highest value of \SI{2.4}{\milli\tesla} at \SI{4}{\kelvin}. 

The dispersion is also compared to micromagnetic simulations, visualized by the colormap. We observe good agreement of the numerical model and the semi-analytical calculations for wavenumbers up to $2000~\mathrm{rad}/\mathrm{cm}$ and an increasing deviation for higher $k$. The dashed lines in Fig.~\ref{fig:fig02}(b) mark the possible wavevectors $k$ that can be excited with the used antenna width of $w_\mathrm{a}~=~\SI{50}{\micro\meter}$, based on a simple $sinc$-shaped excitation efficiency as the Fourier-transform of a rectangular antenna. Figure~\ref{fig:fig02}(c) shows the group velocity for the experimentally accessible wavenumbers, extracted as the derivative of the dispersion relation. The calculated dispersion allows the transformation of the transmission spectra from the frequency domain into the wavenumber domain and compare magnon transport at decreasing temperatures with $k$. 

Figure~\ref{fig:fig02}(d) depicts the experimentally measured $S_{21}$ transmission spectra, transformed into $k$-space with the dispersion relation of Eq.~\ref{Dispersion}. We observe similar behavior of spin-wave magnitude as in the frequency domain, with the smallest $S_{21}$-amplitude and the narrowest spectrum at $\SI{4}{\kelvin}$ and an increase of amplitude and spectrum width again at $\SI{26}{\milli\kelvin}$. However, a comparison of the spectrum width reveals, that almost all wavenumbers, which can be excited by the antenna, are detected at $\SI{293}{\kelvin}$, but less than half of the possible wavenumbers are detected at the cryogenic temperatures. In other words, a faster decrease of spin-wave amplitude with wavenumber was observed at lower temperatures, compared to room temperature.

\begin{figure*}[tb]
  \includegraphics[width=\textwidth]{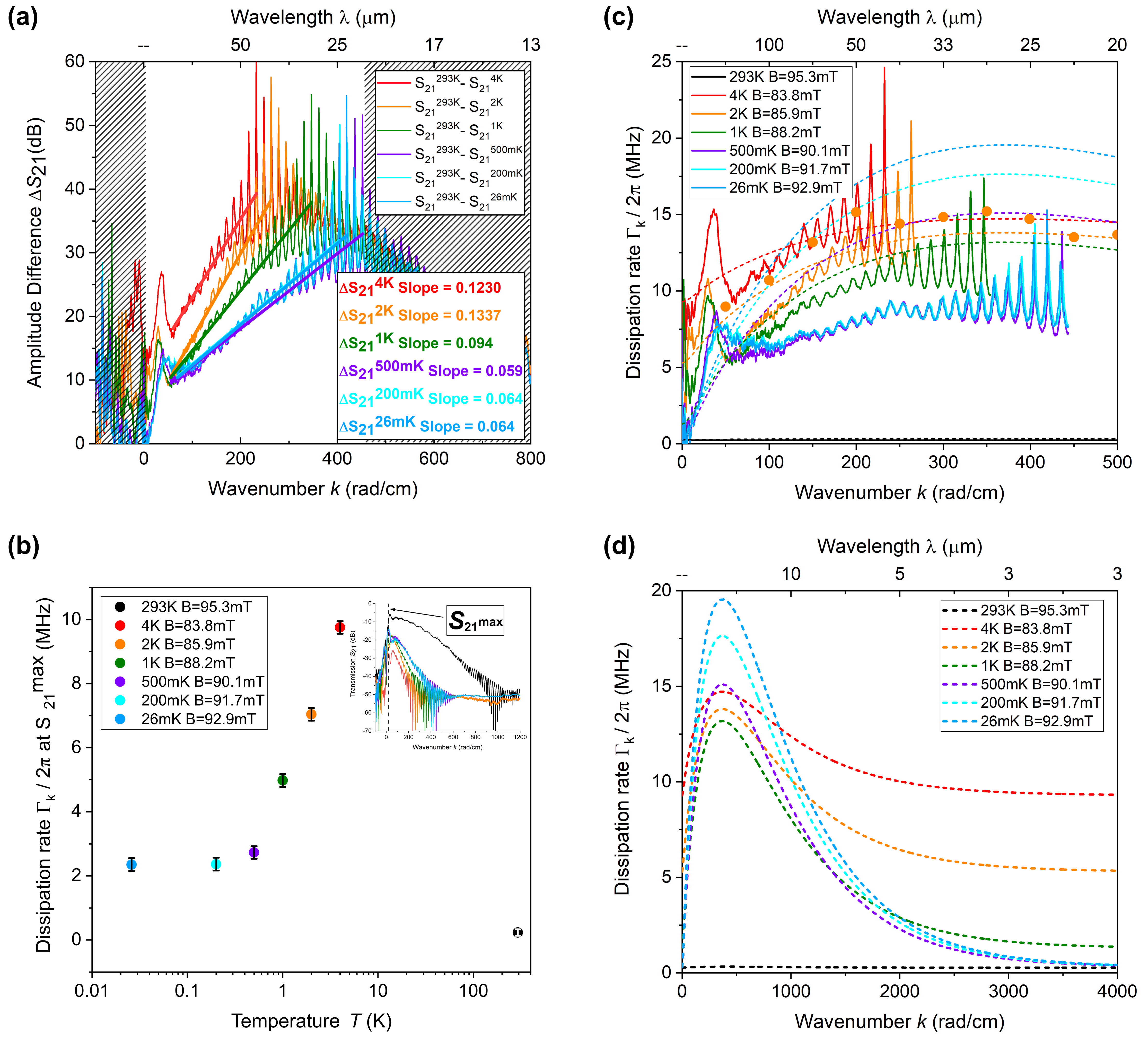}
  \captionsetup{justification=justified}
  \caption{\justifying (a)~Difference in spin-wave amplitude $\Delta S_{21}^{293\mathrm{K - T}}$ with respect to wavenumber. The region of spin-wave propagation was linearly fitted to quantify the change of spin-wave transmission with $k$, compared to room temperature. (b)~Illustration of the temperature dependency of the dissipation rate $\Gamma_\mathrm{k}$, at the point of maximum spin-wave transmission $S_{21}$ in $k$-space. (c)~The dissipation rate $\Gamma_\mathrm{k}$ plotted versus wavenumber in the regions of spin-wave propagation. We compare the results extracted out of the experiment (solid lines), the quasi-analytical model (dashed lines), and the micromagnetic simulations (dots). (d)~Theoretically calculated change of the dissipation rate with wavenumber at different temperatures. After a maximum of dissipation at $400$~rad/cm, the model predicts a steady decrease of $\Gamma_\mathrm{k}$ with increasing wavenumbers.}
  \label{fig:fig04}
\end{figure*}

\subsection{Wavenumber dependency of the dissipation rate}
\label{sec:resultsDEL}
To quantify the observed dependency of magnon transport on the wavenumber, Fig.~\ref{fig:fig04}(a) shows the difference in transmission amplitude $\Delta S_{21}^{293\mathrm{K - T}}$ between room temperature and the cryogenic temperatures. When plotted versus $k$, a linear fit allows to compare the change of $\Delta S_{21}^{293\mathrm{K - T}}$ with wavenumber in the region of spin-wave propagation. The slope of the fit indicates a faster decay of spin waves with $k$ at low temperatures. We observe the steepest slope, and therefore the biggest dependency of magnon transport on $k$, between $\SI{4}{\kelvin}-\SI{2}{\kelvin}$ and a smaller slope at $\SI{500}{\milli\kelvin}$. Below $\SI{500}{\milli\kelvin}$, no notable changes of the amplitude difference $\Delta S_{21}^{293\mathrm{K - T}}$ and its dependency on the wavenumber were recorded. Similar observations have been reported for temperature dependent FMR studies~\cite{Serha2024} and single-crystal magnetometry measurements~\cite{Deen2015}, with no further change of the FMR frequency and the magnetization below \SI{500}{\milli\kelvin}. This unusual behavior is attributed to the complex magnetic phase transition of GGG at temperatures below $\SI{1}{\kelvin}$~\cite{Deen2015, Petrenko1998, Tsui1999, Schiffer1994}.

The magnitude of the transmission $S_{21}$ is a measure for the dissipation rate $\Gamma_\mathrm{k}$, as the inverse of the magnon lifetime~$\tau$, at the respective temperatures (see Eq.~\eqref{S21}). Hence, the experimentally obtained transmission spectra $S_{21}$ allow to calculate $\Gamma_\mathrm{k}$ directly, assuming the efficiency with which the antennas can excite and detect spin waves $J^2 = J_{\mathrm{ex}}\cdot J_{\mathrm{det}}$ is not dependent on temperature. The used microstrip antennas consist of gold plated copper and are patterned on a Rogers RT/duroid 6010.2LM substrate. As both copper and gold show a decrease in resistivity at lower temperatures~\cite{Matula1979} and the substrates dielectric constant $\epsilon_\mathrm{r}=10.2$ does not change significantly with temperature, according to the thermal coefficient of $-425$ ppm/$^\circ$C given by the manufacturer~\cite{datasheetDuroid}, the characteristic impedance, and therefore also the field components of the Oersted field created by the antenna, are assumed to be temperature independent in our experiments. Equation~\eqref{S21} describes the relation between $S_{21}$ (in linear magnitude) and the dissipation rate $\Gamma_\mathrm{k}$, with $L$ as the antenna distance and $v_{\mathrm{g}} = v_{\mathrm{g}}(k)$ the group velocity,
\begin{equation}
    S_{21} = J^2\cdot \exp{\left(-\frac{\Gamma_\mathrm{k} L}{v_\mathrm{g}}\right)}.
    \label{S21}
\end{equation}

The group velocity is extracted as the derivative of the dispersion relation, shown in Fig.~\ref{fig:fig02}(c), $v_{\mathrm{g}}~=~|\partial \omega_\mathrm{k}/\partial k|$. At room temperature conditions, the GGG substrate can be considered to have zero net-magnetization and the MSSW damping rate is $k$-independent and is given by $\Gamma_{293\mathrm{K}}~=~\alpha_{293\mathrm{K}}~\left(\omega_\mathrm{H,YIG}~+~\omega_\mathrm{M,YIG}/2\right)$. In this case, the antenna efficiency $J^2$ can be calculated via Eq.~\eqref{antenna efficiency}, with the room temperature Gilbert damping $\alpha_{293\mathrm{K}}~=~(5.2~\pm~2.5)~\cdot~10^{-5}$, measured experimentally in a standard FMR setup:
\begin{equation}
    J^2 = \frac{S_{21}}{\exp{\left(-\frac{\Gamma_{293\mathrm{K}} L}{v_\mathrm{g}}\right)}}.
    \label{antenna efficiency}
\end{equation}

By assuming $J^2$ is independent of temperature, Eq.~\ref{S21} can be converted for $\Gamma_\mathrm{k}$ (or inverse magnon lifetime), allowing to calculate the dissipation rate at the different temperatures directly out of the recorded PSWS spectra as
\begin{equation}
    \Gamma_\mathrm{k} = \frac{1}{\tau} = -\frac{v_\mathrm{g}\cdot\ln(S_{21}/J^2)}{L}.
    \label{Gamma}
\end{equation}

To better visualize the change of $\Gamma_\mathrm{k}$ with temperature, Fig.~\ref{fig:fig04}(b) compares the dissipation rate for different temperatures at the point of maximum spin-wave transmission in the wavenumber domain, close to the FMR point ($k~=~0$). As expected, the lowest dissipation with $\Gamma_\mathrm{min}/(2\pi)~=~(0.26~\pm~0.1)$~MHz is observed at $\SI{293}{\kelvin}$ and the maximum with $\Gamma_\mathrm{max}/(2\pi~)=~(9.76~\pm~0.20)$~MHz was recorded at $\SI{4}{\kelvin}$. In terms of effective magnetic damping, this would correspond to the values $\alpha_{\mathrm{eff}}^{4\mathrm{K}}~=~(1.7~\pm~0.1)~\cdot~10^{-3}$ and $\alpha_{\mathrm{eff}}^{26\mathrm{mK}}~=~(4.2~\pm~0.5)~\cdot~10^{-4}$. At temperatures below $\SI{4}{\kelvin}$, the dissipation rate decreases continuously, until it reaches a plateau at $\SI{500}{\milli\kelvin}$. This observation aligns with the results regarding magnetic damping from~\cite{Kosen2019} and is related to the different temperature dependent dissipation processes described in section~\ref{sec:resultsSKL}.

Figure~\ref{fig:fig04}(c) depicts the dissipation $\Gamma_\mathrm{k}$ versus wavenumber and demonstrates an evident increase of the damping rate with the wavenumber in almost the whole experimentally accessible range. Certain oscillations are most probably a result of interference of the main MSSW mode with edge modes (which are formed due to inhomogeneous GGG stray fields across the sample width). Analog to Fig.~\ref{fig:fig04}(b), the highest dissipation rate in the experiment was observed at $\SI{4}{\kelvin}$, followed by a continuous decrease of $\Gamma_\mathrm{k}$ until $\SI{500}{\milli\kelvin}$, from where no significant changes can be observed by decreasing the temperature further. The dissipation rate increased with wavenumber by approximately $70\%$ at \SI{4}{\kelvin} and by $55\%$ at \SI{26}{\milli\kelvin} between the FMR-point and the highest excited $k$, indicating that the dependency of $\Gamma_\mathrm{k}$ on the wavenumber becomes smaller at sub-Kelvin temperatures. The above-mentioned mechanism of rare-earth ion relaxations in YIG is expected to be $k$-independent and results in a uniform increase of the dissipation rate, at least in the studied wavenumber range (the behavior might differ for much shorter spin-wave wavelengths, comparable to the mean distance between the relaxation centers). In order to understand the experimental data, we compare them with the results of the semi-analytical calculations and the micromagnetic simulations, which are shown by dashed lines and dots in Fig.~\ref{fig:fig04}(c). From the micromagnetic simulations, the dissipation was only extracted at $\SI{2}{\kelvin}$, due to computational resources.  

Both the analytical and micromagnetic results show a pronounced increase of the dissipation rate $\Gamma_\mathrm{k}$ with the wavenumber in the range of experimentally recorded magnons, and are in agreement with each other, see Fig.~\ref{fig:fig04}(c). These models do not account for the rare-earth ion relaxation, but allow to capture the influence of the partially magnetized paramagnetic GGG substrate on the MSSW relaxation, mediated by the dynamic dipolar coupling between the YIG and GGG layer. This dipolar coupling is the source of the observed $k$-dependence of $\Gamma_\mathrm{k}$. Indeed, the intrinsic damping rate in GGG is much larger than in YIG. Like in the framework of classical coupled harmonic oscillators, the effective damping rate of higher-Q oscillators (lower inherent damping) increases due to the energy transfer into lower-Q oscillators (higher inherent damping). As we deal in our case with propagating waves instead of uniform oscillations, the energy transfer and additional dissipation rate is dependent on the wavenumber, since the dipolar coupling is $k$-dependent. This coupling is described by the magnetostatic Green function, which nonzero components in the film geometry are expressed as \cite{Kalinikos1986}
    \begin{equation}\label{e:Gk}
    \begin{split}
        G_\mathrm{k,xx} &= - G_\mathrm{k,zz} = \frac{|k_\mathrm{x}|}{2} e^{-|k_\mathrm{x} (z-z')|}\,, \\ 
        G_\mathrm{k,xz} &= G_\mathrm{k,zx} = i\frac{k_\mathrm{x}}{2} \mathrm{sign}[z-z']  e^{-|k_\mathrm{x} (z-z')|}\,,
    \end{split}
    \end{equation}
where the wave propagates along $x$-direction and $z$ and $z'$ are perpendicular to the plane coordinates within different layers (the total coupling is an integral over the layers thickness).  At the FMR point ($k_\mathrm{x} = 0$), the dipolar coupling between YIG and the partially magnetized GGG is zero, as expected for the coupling of uniformly magnetized films. Then, at low $k$, the coupling increases linearly with the wavenumber, leading to the increase of additional losses of MSSW in YIG. 

Analytical calculations also revealed that the GGG-induced damping $\Gamma^\mathrm{(dip)}$ is added linearly to the intrinsic YIG damping, $\Gamma_\mathrm{k} = \Gamma_0 + \Gamma^\mathrm{(dip)}$. This feature is used in Fig.~\ref{fig:fig04}(c) for the fitting of the experimental data with the semi-analytical model of Eq.~\ref{Dispersion}, where we use the temperature-dependent $\Gamma_0$ as a fitting parameter. By including the obtained additional YIG inherent dissipation due to rare-earth ion relaxations $\Gamma_0(\SI{4}{\kelvin})~=~\SI{9}{\mega\hertz}$, $\Gamma_0(\SI{2}{\kelvin})~=~\SI{5}{\mega\hertz}$, and $\Gamma_0(\SI{1}{\kelvin})~=~\SI{1}{\mega\hertz}$, the theoretical curves get into good agreement with the experimental data between $4$ to $\SI{1}{\kelvin}$. Below \SI{1}{\kelvin} however, the theoretical predictions deviate significantly from the experiment and overestimate the change of the dissipation rate $\Gamma_\mathrm{k}$ with wavenumber. As the theoretical model does not consider the complex phase transitions of the paramagnetic GGG substrate, it predicts a continuous increase of the dissipation between \SI{500}{\milli\kelvin} and \SI{26}{\milli\kelvin}, due to the increased GGG net magnetization (within the Brillouin model of paramagnetic GGG \cite{Barak1992, Serha2024}). This contradicts the dissipation in the experiment, which steadily decreases down to $\SI{500}{\milli\kelvin}$ and remains almost unchanged at lower temperatures. To include the complex magnetic nature of GGG into the calculations, further temperature-dependent studies of the linewidth of pure GGG samples are needed at sub-kelvin temperatures.

As the signal-to-noise ratio at cryogenic temperatures was already too small to resolve spin-wave transmission at higher wavenumbers, we were only able to experimentally record spin-wave propagation for wavenumbers up to $450$~rad/cm. To illustrate the behavior of $\Gamma_\mathrm{k}$ for short-wavelength spin waves, Fig.~\ref{fig:fig04}(d) depicts the dissipation rate computed with Eq.~\ref{Dispersion} for wavenumbers up to $4000$~rad/cm. Note, that the quasi-analytical model shows a steady decrease of the dissipation $\Gamma_\mathrm{k}$ at wavenumbers $k~>~400$~rad/cm for the whole temperature range of \SI{4}{\kelvin} to \SI{26}{\milli\kelvin}. The presented theoretical model demonstrates that the increased dissipation rate, introduced by the dipolar coupling between the YIG layer and the partially magnetized GGG substrate, is suppressed at increasing wavenumbers. As Eq.~\eqref{e:Gk} shows, at high enough $k$ (quantitatively, above $k_\mathrm{x} d_\mathrm{YIG} \gtrsim 3$), the exponential decrease dominates over the linear increase with wavenumber.  Hence, due to the demand of low magnetic losses in the field of quantum magnonics, it could proof beneficial to work with short-wavelength exchange-dominated magnons in future experiments.

\section{Conclusions}
\label{sec:conclusion}
In conclusion, we have shown magnon transport at different temperature ranges, from $\SI{293}{\kelvin}$ to $\SI{26}{\milli\kelvin}$, in a $\SI{7.78}{\micro\meter}$-thick YIG sample. The recorded $S_{21}$ transmission spectra were converted into the wavenumber domain via the semi-analytically calculated dispersion relation for a two-layered magnetic structure and confirmed via micromagnetic simulations. Both the theoretical as well as the numerical calculations considered the stray field of the partially magnetized GGG substrate at decreasing temperatures. Furthermore, the experimentally obtained $S_{21}$-spectra were directly converted to the dissipation rate $\Gamma_\mathrm{k}$ via the relation of Eq.~\ref{Gamma}. At cryogenic temperatures, we report a distinct dependency of $\Gamma_\mathrm{k}$ on the wavenumber, with an increase of up to $55\%$ between the FMR point and the highest excited wavenumber at the temperature of \SI{26}{\milli\kelvin}. Furthermore, the theoretical model predicts a steady decrease of dissipation at wavenumbers $k>400$~rad/cm, suggesting lower dissipation for short-wavelength exchange magnons. We attribute the recorded $k$-dependency of $\Gamma_\mathrm{k}$ to the different magnetic resonance conditions of the dipolar coupled ferrimagnetic YIG layer and the partially magnetized GGG substrate, which is suppressed at higher $k$ due to the smaller dipolar coupling. Additionally, our experiments confirmed an increase in magnetic damping due to rare-earth ion relaxations at low temperatures and the subsequent decrease of damping at sub-kelvin temperatures, which is already reported in the literature. Below \SI{500}{\milli\kelvin}, we recorded no changes in spin-wave propagation or the dissipation rate, due to the complex phase diagram of GGG below temperatures of \SI{1}{\kelvin}. The coupling between YIG and the necessary GGG substrate at decreasing temperatures and the concomitant increase in magnetic losses should be taken into account in the investigation of magnons on the quantum level, especially with propagating, rather than standing, spin-waves.\\

\section*{Acknowledgements}
\label{sec:acknowledgements}
Work has been supported by the Austrian Science Fund FWF in the frame of the project Paramagnonics (10.55776/I6568). R.V. and DE.S. acknowledge the support of the NAS of Ukraine (projects 0123U104827 and 0123U100898). S.K. acknowledges the support by the H2020-MSCA-IF under Grant No. 101025758 (“OMNI”). K.O.L. acknowledges the Austrian Science Fund FWF for the support through ESPRIT Fellowship Grant TopMag (10.55776/ESP526). C.A. acknowledges the support by the FWF project 10.55776/I6068. The computational results presented have been achieved using the Vienna Scientific Cluster (VSC). The authors would like to thank Barbora and Sabri Koraltan for their wedding, resulting in valuable discussions.

DA.S. conducted all measurements, processed and analyzed the data, and authored the initial draft of the manuscript. A.A.V. established and analyzed the micromagnetic simulations. R.O.S. assisted in analyzing and interpreting the experimental and theoretical results. R.V. and DE.S. established the semi-analytical dispersion calculation and supported the analysis and interpretation of the results. S.K., DA.S., R.O.S., and K.O.L. constructed the experimental setup. DI.S. and C.A. developed the software for the micromagnetic simulations and supervised the simulations. A.V.C planned the experiment and led the project. All authors discussed the results and contributed to the manuscript. 

The authors declare no competing interests.

\bibliography{bibliography}  

\begin{thebibliography}{55}%
\makeatletter
\providecommand \@ifxundefined [1]{%
 \@ifx{#1\undefined}
}%
\providecommand \@ifnum [1]{%
 \ifnum #1\expandafter \@firstoftwo
 \else \expandafter \@secondoftwo
 \fi
}%
\providecommand \@ifx [1]{%
 \ifx #1\expandafter \@firstoftwo
 \else \expandafter \@secondoftwo
 \fi
}%
\providecommand \natexlab [1]{#1}%
\providecommand \enquote  [1]{``#1''}%
\providecommand \bibnamefont  [1]{#1}%
\providecommand \bibfnamefont [1]{#1}%
\providecommand \citenamefont [1]{#1}%
\providecommand \href@noop [0]{\@secondoftwo}%
\providecommand \href [0]{\begingroup \@sanitize@url \@href}%
\providecommand \@href[1]{\@@startlink{#1}\@@href}%
\providecommand \@@href[1]{\endgroup#1\@@endlink}%
\providecommand \@sanitize@url [0]{\catcode `\\12\catcode `\$12\catcode `\&12\catcode `\#12\catcode `\^12\catcode `\_12\catcode `\%12\relax}%
\providecommand \@@startlink[1]{}%
\providecommand \@@endlink[0]{}%
\providecommand \url  [0]{\begingroup\@sanitize@url \@url }%
\providecommand \@url [1]{\endgroup\@href {#1}{\urlprefix }}%
\providecommand \urlprefix  [0]{URL }%
\providecommand \Eprint [0]{\href }%
\providecommand \doibase [0]{https://doi.org/}%
\providecommand \selectlanguage [0]{\@gobble}%
\providecommand \bibinfo  [0]{\@secondoftwo}%
\providecommand \bibfield  [0]{\@secondoftwo}%
\providecommand \translation [1]{[#1]}%
\providecommand \BibitemOpen [0]{}%
\providecommand \bibitemStop [0]{}%
\providecommand \bibitemNoStop [0]{.\EOS\space}%
\providecommand \EOS [0]{\spacefactor3000\relax}%
\providecommand \BibitemShut  [1]{\csname bibitem#1\endcsname}%
\let\auto@bib@innerbib\@empty
\bibitem [{\citenamefont {Serga}\ \emph {et~al.}(2010)\citenamefont {Serga}, \citenamefont {Chumak},\ and\ \citenamefont {Hillebrands}}]{Serga2010}%
  \BibitemOpen
  \bibfield  {author} {\bibinfo {author} {\bibfnamefont {A.~A.}\ \bibnamefont {Serga}}, \bibinfo {author} {\bibfnamefont {A.~V.}\ \bibnamefont {Chumak}},\ and\ \bibinfo {author} {\bibfnamefont {B.}~\bibnamefont {Hillebrands}},\ }\bibfield  {title} {\bibinfo {title} {Yig magnonics},\ }\href {https://doi.org/10.1088/0022-3727/43/26/264002} {\bibfield  {journal} {\bibinfo  {journal} {J. Phys. D: Appl. Phys.}\ }\textbf {\bibinfo {volume} {43}},\ \bibinfo {pages} {264002} (\bibinfo {year} {2010})}\BibitemShut {NoStop}%
\bibitem [{\citenamefont {Adam}(1988)}]{Adam1988}%
  \BibitemOpen
  \bibfield  {author} {\bibinfo {author} {\bibfnamefont {J.}~\bibnamefont {Adam}},\ }\bibfield  {title} {\bibinfo {title} {Analog signal processing with microwave magnetics},\ }\href {https://doi.org/10.1109/5.4392} {\bibfield  {journal} {\bibinfo  {journal} {Proc. IEEE}\ }\textbf {\bibinfo {volume} {76}},\ \bibinfo {pages} {159} (\bibinfo {year} {1988})}\BibitemShut {NoStop}%
\bibitem [{\citenamefont {Glass}(1988)}]{Glass1988}%
  \BibitemOpen
  \bibfield  {author} {\bibinfo {author} {\bibfnamefont {H.}~\bibnamefont {Glass}},\ }\bibfield  {title} {\bibinfo {title} {Ferrite films for microwave and millimeter-wave devices},\ }\href {https://doi.org/10.1109/5.4391} {\bibfield  {journal} {\bibinfo  {journal} {Proc. IEEE}\ }\textbf {\bibinfo {volume} {76}},\ \bibinfo {pages} {151} (\bibinfo {year} {1988})}\BibitemShut {NoStop}%
\bibitem [{\citenamefont {Ishak}(1988)}]{Ishak1988}%
  \BibitemOpen
  \bibfield  {author} {\bibinfo {author} {\bibfnamefont {W.}~\bibnamefont {Ishak}},\ }\bibfield  {title} {\bibinfo {title} {Magnetostatic wave technology: a review},\ }\href {https://doi.org/10.1109/5.4393} {\bibfield  {journal} {\bibinfo  {journal} {Proc. IEEE}\ }\textbf {\bibinfo {volume} {76}},\ \bibinfo {pages} {171} (\bibinfo {year} {1988})}\BibitemShut {NoStop}%
\bibitem [{\citenamefont {Morgenthaler}(1988)}]{Morgenthaler1988}%
  \BibitemOpen
  \bibfield  {author} {\bibinfo {author} {\bibfnamefont {F.}~\bibnamefont {Morgenthaler}},\ }\bibfield  {title} {\bibinfo {title} {An overview of electromagnetic and spin angular momentum mechanical waves in ferrite media},\ }\href {https://doi.org/10.1109/5.4390} {\bibfield  {journal} {\bibinfo  {journal} {Proc. IEEE}\ }\textbf {\bibinfo {volume} {76}},\ \bibinfo {pages} {138} (\bibinfo {year} {1988})}\BibitemShut {NoStop}%
\bibitem [{\citenamefont {Rodrigue}(1988)}]{Rodrigue1988}%
  \BibitemOpen
  \bibfield  {author} {\bibinfo {author} {\bibfnamefont {G.}~\bibnamefont {Rodrigue}},\ }\bibfield  {title} {\bibinfo {title} {A generation of microwave ferrite devices},\ }\href {https://doi.org/10.1109/5.4389} {\bibfield  {journal} {\bibinfo  {journal} {Proc. IEEE}\ }\textbf {\bibinfo {volume} {76}},\ \bibinfo {pages} {121} (\bibinfo {year} {1988})}\BibitemShut {NoStop}%
\bibitem [{\citenamefont {Zenbaa}\ \emph {et~al.}()\citenamefont {Zenbaa}, \citenamefont {Abert}, \citenamefont {Majcen}, \citenamefont {Kerber}, \citenamefont {Serha}, \citenamefont {Knauer}, \citenamefont {Wang}, \citenamefont {Schrefl}, \citenamefont {Suess},\ and\ \citenamefont {Chumak}}]{Zeenba2024}%
  \BibitemOpen
  \bibfield  {author} {\bibinfo {author} {\bibfnamefont {N.}~\bibnamefont {Zenbaa}}, \bibinfo {author} {\bibfnamefont {C.}~\bibnamefont {Abert}}, \bibinfo {author} {\bibfnamefont {F.}~\bibnamefont {Majcen}}, \bibinfo {author} {\bibfnamefont {M.}~\bibnamefont {Kerber}}, \bibinfo {author} {\bibfnamefont {R.~O.}\ \bibnamefont {Serha}}, \bibinfo {author} {\bibfnamefont {S.}~\bibnamefont {Knauer}}, \bibinfo {author} {\bibfnamefont {Q.}~\bibnamefont {Wang}}, \bibinfo {author} {\bibfnamefont {T.}~\bibnamefont {Schrefl}}, \bibinfo {author} {\bibfnamefont {D.}~\bibnamefont {Suess}},\ and\ \bibinfo {author} {\bibfnamefont {A.~V.}\ \bibnamefont {Chumak}},\ }\bibfield  {title} {\bibinfo {title} {Experimental realisation of a universal inverse-design magnonic device},\ }\href {http://arxiv.org/abs/2403.17724} {\bibfield  {journal} {\bibinfo  {journal} {arXiv}\ }}\Eprint {https://arxiv.org/abs/2403.17724} {arXiv:2403.17724} \BibitemShut {NoStop}%
\bibitem [{\citenamefont {Finocchio}\ \emph {et~al.}(2024)\citenamefont {Finocchio} \emph {et~al.}}]{Finocchio2024}%
  \BibitemOpen
  \bibfield  {author} {\bibinfo {author} {\bibfnamefont {G.}~\bibnamefont {Finocchio}} \emph {et~al.},\ }\bibfield  {title} {\bibinfo {title} {Roadmap for unconventional computing with nanotechnology},\ }\href {https://doi.org/10.1088/2399-1984/ad299a} {\bibfield  {journal} {\bibinfo  {journal} {Nano Futures}\ }\textbf {\bibinfo {volume} {8}},\ \bibinfo {eid} {012001} (\bibinfo {year} {2024})}\BibitemShut {NoStop}%
\bibitem [{\citenamefont {Wang}\ \emph {et~al.}(2024{\natexlab{a}})\citenamefont {Wang}, \citenamefont {Csaba}, \citenamefont {Verba}, \citenamefont {Chumak},\ and\ \citenamefont {Pirro}}]{Wang2024}%
  \BibitemOpen
  \bibfield  {author} {\bibinfo {author} {\bibfnamefont {Q.}~\bibnamefont {Wang}}, \bibinfo {author} {\bibfnamefont {G.}~\bibnamefont {Csaba}}, \bibinfo {author} {\bibfnamefont {R.}~\bibnamefont {Verba}}, \bibinfo {author} {\bibfnamefont {A.~V.}\ \bibnamefont {Chumak}},\ and\ \bibinfo {author} {\bibfnamefont {P.}~\bibnamefont {Pirro}},\ }\bibfield  {title} {\bibinfo {title} {Nanoscale magnonic networks},\ }\href {https://doi.org/10.1103/PhysRevApplied.21.040503} {\bibfield  {journal} {\bibinfo  {journal} {Phys. Rev. Appl.}\ }\textbf {\bibinfo {volume} {21}},\ \bibinfo {eid} {040503} (\bibinfo {year} {2024}{\natexlab{a}})}\BibitemShut {NoStop}%
\bibitem [{\citenamefont {Wang}\ \emph {et~al.}(2023)\citenamefont {Wang}, \citenamefont {Verba}, \citenamefont {Heinz}, \citenamefont {Schneider}, \citenamefont {Wojewoda}, \citenamefont {Davídková}, \citenamefont {Levchenko}, \citenamefont {Dubs}, \citenamefont {Mauser}, \citenamefont {Urbánek}, \citenamefont {Pirro},\ and\ \citenamefont {Chumak}}]{Wang2023}%
  \BibitemOpen
  \bibfield  {author} {\bibinfo {author} {\bibfnamefont {Q.}~\bibnamefont {Wang}}, \bibinfo {author} {\bibfnamefont {R.}~\bibnamefont {Verba}}, \bibinfo {author} {\bibfnamefont {B.}~\bibnamefont {Heinz}}, \bibinfo {author} {\bibfnamefont {M.}~\bibnamefont {Schneider}}, \bibinfo {author} {\bibfnamefont {O.}~\bibnamefont {Wojewoda}}, \bibinfo {author} {\bibfnamefont {K.}~\bibnamefont {Davídková}}, \bibinfo {author} {\bibfnamefont {K.}~\bibnamefont {Levchenko}}, \bibinfo {author} {\bibfnamefont {C.}~\bibnamefont {Dubs}}, \bibinfo {author} {\bibfnamefont {N.~J.}\ \bibnamefont {Mauser}}, \bibinfo {author} {\bibfnamefont {M.}~\bibnamefont {Urbánek}}, \bibinfo {author} {\bibfnamefont {P.}~\bibnamefont {Pirro}},\ and\ \bibinfo {author} {\bibfnamefont {A.~V.}\ \bibnamefont {Chumak}},\ }\bibfield  {title} {\bibinfo {title} {Deeply nonlinear excitation of self-normalized short spin waves},\ }\href {https://doi.org/10.1126/sciadv.adg4609} {\bibfield  {journal} {\bibinfo  {journal} {Sci. Adv}\ }\textbf {\bibinfo
  {volume} {9}},\ \bibinfo {eid} {eadg4609} (\bibinfo {year} {2023})}\BibitemShut {NoStop}%
\bibitem [{\citenamefont {Pirro}\ \emph {et~al.}(2021)\citenamefont {Pirro}, \citenamefont {Vasyuchka}, \citenamefont {Serga},\ and\ \citenamefont {Hillebrands}}]{Pirro2021}%
  \BibitemOpen
  \bibfield  {author} {\bibinfo {author} {\bibfnamefont {P.}~\bibnamefont {Pirro}}, \bibinfo {author} {\bibfnamefont {V.~I.}\ \bibnamefont {Vasyuchka}}, \bibinfo {author} {\bibfnamefont {A.~A.}\ \bibnamefont {Serga}},\ and\ \bibinfo {author} {\bibfnamefont {B.}~\bibnamefont {Hillebrands}},\ }\bibfield  {title} {\bibinfo {title} {Advances in coherent magnonics},\ }\href {https://doi.org/10.1038/s41578-021-00332-w} {\bibfield  {journal} {\bibinfo  {journal} {Nat. Rev. Mater.}\ }\textbf {\bibinfo {volume} {6}},\ \bibinfo {pages} {1114} (\bibinfo {year} {2021})}\BibitemShut {NoStop}%
\bibitem [{\citenamefont {Mahmoud}\ \emph {et~al.}(2020)\citenamefont {Mahmoud}, \citenamefont {Ciubotaru}, \citenamefont {Vanderveken}, \citenamefont {Chumak}, \citenamefont {Hamdioui}, \citenamefont {Adelmann},\ and\ \citenamefont {Cotofana}}]{Mahmoud2020}%
  \BibitemOpen
  \bibfield  {author} {\bibinfo {author} {\bibfnamefont {A.}~\bibnamefont {Mahmoud}}, \bibinfo {author} {\bibfnamefont {F.}~\bibnamefont {Ciubotaru}}, \bibinfo {author} {\bibfnamefont {F.}~\bibnamefont {Vanderveken}}, \bibinfo {author} {\bibfnamefont {A.~V.}\ \bibnamefont {Chumak}}, \bibinfo {author} {\bibfnamefont {S.}~\bibnamefont {Hamdioui}}, \bibinfo {author} {\bibfnamefont {C.}~\bibnamefont {Adelmann}},\ and\ \bibinfo {author} {\bibfnamefont {S.}~\bibnamefont {Cotofana}},\ }\bibfield  {title} {\bibinfo {title} {Introduction to spin wave computing},\ }\href {https://doi.org/10.1063/5.0019328} {\bibfield  {journal} {\bibinfo  {journal} {J. Appl. Phys.}\ }\textbf {\bibinfo {volume} {128}},\ \bibinfo {eid} {161101} (\bibinfo {year} {2020})}\BibitemShut {NoStop}%
\bibitem [{\citenamefont {Heinz}\ \emph {et~al.}(2020)\citenamefont {Heinz}, \citenamefont {Brächer}, \citenamefont {Schneider}, \citenamefont {Wang}, \citenamefont {Lägel}, \citenamefont {Friedel}, \citenamefont {Breitbach}, \citenamefont {Steinert}, \citenamefont {Meyer}, \citenamefont {Kewenig}, \citenamefont {Dubs}, \citenamefont {Pirro},\ and\ \citenamefont {Chumak}}]{Heinz2020}%
  \BibitemOpen
  \bibfield  {author} {\bibinfo {author} {\bibfnamefont {B.}~\bibnamefont {Heinz}}, \bibinfo {author} {\bibfnamefont {T.}~\bibnamefont {Brächer}}, \bibinfo {author} {\bibfnamefont {M.}~\bibnamefont {Schneider}}, \bibinfo {author} {\bibfnamefont {Q.}~\bibnamefont {Wang}}, \bibinfo {author} {\bibfnamefont {B.}~\bibnamefont {Lägel}}, \bibinfo {author} {\bibfnamefont {A.~M.}\ \bibnamefont {Friedel}}, \bibinfo {author} {\bibfnamefont {D.}~\bibnamefont {Breitbach}}, \bibinfo {author} {\bibfnamefont {S.}~\bibnamefont {Steinert}}, \bibinfo {author} {\bibfnamefont {T.}~\bibnamefont {Meyer}}, \bibinfo {author} {\bibfnamefont {M.}~\bibnamefont {Kewenig}}, \bibinfo {author} {\bibfnamefont {C.}~\bibnamefont {Dubs}}, \bibinfo {author} {\bibfnamefont {P.}~\bibnamefont {Pirro}},\ and\ \bibinfo {author} {\bibfnamefont {A.~V.}\ \bibnamefont {Chumak}},\ }\bibfield  {title} {\bibinfo {title} {Propagation of spin-wave packets in individual nanosized yttrium iron garnet magnonic conduits},\ }\href
  {https://doi.org/10.1021/acs.nanolett.0c00657} {\bibfield  {journal} {\bibinfo  {journal} {Nano Lett.}\ }\textbf {\bibinfo {volume} {20}},\ \bibinfo {pages} {4220} (\bibinfo {year} {2020})}\BibitemShut {NoStop}%
\bibitem [{\citenamefont {Wang}\ \emph {et~al.}(2024{\natexlab{b}})\citenamefont {Wang}, \citenamefont {Verba}, \citenamefont {Davídková}, \citenamefont {Heinz}, \citenamefont {Tian}, \citenamefont {Rao}, \citenamefont {Guo}, \citenamefont {Guo}, \citenamefont {Dubs}, \citenamefont {Pirro},\ and\ \citenamefont {Chumak}}]{WangNature}%
  \BibitemOpen
  \bibfield  {author} {\bibinfo {author} {\bibfnamefont {Q.}~\bibnamefont {Wang}}, \bibinfo {author} {\bibfnamefont {R.}~\bibnamefont {Verba}}, \bibinfo {author} {\bibfnamefont {K.}~\bibnamefont {Davídková}}, \bibinfo {author} {\bibfnamefont {B.}~\bibnamefont {Heinz}}, \bibinfo {author} {\bibfnamefont {S.}~\bibnamefont {Tian}}, \bibinfo {author} {\bibfnamefont {Y.}~\bibnamefont {Rao}}, \bibinfo {author} {\bibfnamefont {M.}~\bibnamefont {Guo}}, \bibinfo {author} {\bibfnamefont {X.}~\bibnamefont {Guo}}, \bibinfo {author} {\bibfnamefont {C.}~\bibnamefont {Dubs}}, \bibinfo {author} {\bibfnamefont {P.}~\bibnamefont {Pirro}},\ and\ \bibinfo {author} {\bibfnamefont {A.~V.}\ \bibnamefont {Chumak}},\ }\bibfield  {title} {\bibinfo {title} {All-magnonic repeater based on bistability},\ }\href {https://doi.org/10.1038/s41467-024-52084-0} {\bibfield  {journal} {\bibinfo  {journal} {Nat. Commun.}\ }\textbf {\bibinfo {volume} {15}},\ \bibinfo {eid} {7577} (\bibinfo {year} {2024}{\natexlab{b}})}\BibitemShut {NoStop}%
\bibitem [{\citenamefont {Heussner}\ \emph {et~al.}(2020)\citenamefont {Heussner}, \citenamefont {Talmelli}, \citenamefont {Geilen}, \citenamefont {Heinz}, \citenamefont {Brächer}, \citenamefont {Meyer}, \citenamefont {Ciubotaru}, \citenamefont {Adelmann}, \citenamefont {Yamamoto}, \citenamefont {Serga}, \citenamefont {Hillebrands},\ and\ \citenamefont {Pirro}}]{Heussner2020}%
  \BibitemOpen
  \bibfield  {author} {\bibinfo {author} {\bibfnamefont {F.}~\bibnamefont {Heussner}}, \bibinfo {author} {\bibfnamefont {G.}~\bibnamefont {Talmelli}}, \bibinfo {author} {\bibfnamefont {M.}~\bibnamefont {Geilen}}, \bibinfo {author} {\bibfnamefont {B.}~\bibnamefont {Heinz}}, \bibinfo {author} {\bibfnamefont {T.}~\bibnamefont {Brächer}}, \bibinfo {author} {\bibfnamefont {T.}~\bibnamefont {Meyer}}, \bibinfo {author} {\bibfnamefont {F.}~\bibnamefont {Ciubotaru}}, \bibinfo {author} {\bibfnamefont {C.}~\bibnamefont {Adelmann}}, \bibinfo {author} {\bibfnamefont {K.}~\bibnamefont {Yamamoto}}, \bibinfo {author} {\bibfnamefont {A.~A.}\ \bibnamefont {Serga}}, \bibinfo {author} {\bibfnamefont {B.}~\bibnamefont {Hillebrands}},\ and\ \bibinfo {author} {\bibfnamefont {P.}~\bibnamefont {Pirro}},\ }\bibfield  {title} {\bibinfo {title} {Experimental realization of a passive gigahertz frequency-division demultiplexer for magnonic logic networks},\ }\href {https://doi.org/10.1002/pssr.201900695} {\bibfield  {journal} {\bibinfo
  {journal} {Phys. Status Solidi RRL}\ }\textbf {\bibinfo {volume} {14}},\ \bibinfo {eid} {1900695} (\bibinfo {year} {2020})}\BibitemShut {NoStop}%
\bibitem [{\citenamefont {Vogt}\ \emph {et~al.}(2014)\citenamefont {Vogt}, \citenamefont {Fradin}, \citenamefont {Pearson}, \citenamefont {Sebastian}, \citenamefont {Bader}, \citenamefont {Hillebrands}, \citenamefont {Hoffmann},\ and\ \citenamefont {Schultheiss}}]{Vogt2014}%
  \BibitemOpen
  \bibfield  {author} {\bibinfo {author} {\bibfnamefont {K.}~\bibnamefont {Vogt}}, \bibinfo {author} {\bibfnamefont {F.~Y.}\ \bibnamefont {Fradin}}, \bibinfo {author} {\bibfnamefont {J.~E.}\ \bibnamefont {Pearson}}, \bibinfo {author} {\bibfnamefont {T.}~\bibnamefont {Sebastian}}, \bibinfo {author} {\bibfnamefont {S.~D.}\ \bibnamefont {Bader}}, \bibinfo {author} {\bibfnamefont {B.}~\bibnamefont {Hillebrands}}, \bibinfo {author} {\bibfnamefont {A.}~\bibnamefont {Hoffmann}},\ and\ \bibinfo {author} {\bibfnamefont {H.}~\bibnamefont {Schultheiss}},\ }\bibfield  {title} {\bibinfo {title} {Realization of a spin-wave multiplexer},\ }\href {https://doi.org/10.1038/ncomms4727} {\bibfield  {journal} {\bibinfo  {journal} {Nat. Commun.}\ }\textbf {\bibinfo {volume} {5}},\ \bibinfo {eid} {3727} (\bibinfo {year} {2014})}\BibitemShut {NoStop}%
\bibitem [{\citenamefont {Zavislyak}\ and\ \citenamefont {Popov}(2011)}]{Yttrium2009}%
  \BibitemOpen
  \bibfield  {author} {\bibinfo {author} {\bibfnamefont {I.}~\bibnamefont {Zavislyak}}\ and\ \bibinfo {author} {\bibfnamefont {M.}~\bibnamefont {Popov}},\ }in\ \href@noop {} {\emph {\bibinfo {booktitle} {Yttrium: Compounds, Production and Applications}}},\ \bibinfo {editor} {edited by\ \bibinfo {editor} {\bibfnamefont {B.~D.}\ \bibnamefont {Volkerts}}}\ (\bibinfo  {publisher} {Nova Science Publishers, Inc},\ \bibinfo {address} {New York, USA},\ \bibinfo {year} {2011})\ Chap.~\bibinfo {chapter} {3}, pp.\ \bibinfo {pages} {87--125}\BibitemShut {NoStop}%
\bibitem [{\citenamefont {Chumak}\ \emph {et~al.}(2022)\citenamefont {Chumak} \emph {et~al.}}]{Chumak2022}%
  \BibitemOpen
  \bibfield  {author} {\bibinfo {author} {\bibfnamefont {A.~V.}\ \bibnamefont {Chumak}} \emph {et~al.},\ }\bibfield  {title} {\bibinfo {title} {Advances in magnetics roadmap on spin-wave computing},\ }\href {https://doi.org/10.1109/TMAG.2022.3149664} {\bibfield  {journal} {\bibinfo  {journal} {IEEE Trans. Magn.}\ }\textbf {\bibinfo {volume} {58}},\ \bibinfo {eid} {0800172} (\bibinfo {year} {2022})}\BibitemShut {NoStop}%
\bibitem [{\citenamefont {Barman}\ \emph {et~al.}(2021)\citenamefont {Barman} \emph {et~al.}}]{Barman2021}%
  \BibitemOpen
  \bibfield  {author} {\bibinfo {author} {\bibfnamefont {A.}~\bibnamefont {Barman}} \emph {et~al.},\ }\bibfield  {title} {\bibinfo {title} {The 2021 magnonics roadmap},\ }\href {https://doi.org/10.1088/1361-648X/abec1a} {\bibfield  {journal} {\bibinfo  {journal} {J. Phys.: Condens.Matter}\ }\textbf {\bibinfo {volume} {33}},\ \bibinfo {eid} {413001} (\bibinfo {year} {2021})}\BibitemShut {NoStop}%
\bibitem [{\citenamefont {Zhang}(2023)}]{Zhang2023}%
  \BibitemOpen
  \bibfield  {author} {\bibinfo {author} {\bibfnamefont {X.}~\bibnamefont {Zhang}},\ }\bibfield  {title} {\bibinfo {title} {A review of common materials for hybrid quantum magnonics},\ }\href {https://doi.org/10.1016/j.mtelec.2023.100044} {\bibfield  {journal} {\bibinfo  {journal} {Materials Today Electronics}\ }\textbf {\bibinfo {volume} {5}},\ \bibinfo {eid} {100044} (\bibinfo {year} {2023})}\BibitemShut {NoStop}%
\bibitem [{\citenamefont {Jiang}\ \emph {et~al.}(2023)\citenamefont {Jiang}, \citenamefont {Lim}, \citenamefont {Li}, \citenamefont {Pfaff}, \citenamefont {Lo}, \citenamefont {Qian}, \citenamefont {Schleife}, \citenamefont {Zuo}, \citenamefont {Novosad},\ and\ \citenamefont {Hoffmann}}]{Jiang2023}%
  \BibitemOpen
  \bibfield  {author} {\bibinfo {author} {\bibfnamefont {Z.}~\bibnamefont {Jiang}}, \bibinfo {author} {\bibfnamefont {J.}~\bibnamefont {Lim}}, \bibinfo {author} {\bibfnamefont {Y.}~\bibnamefont {Li}}, \bibinfo {author} {\bibfnamefont {W.}~\bibnamefont {Pfaff}}, \bibinfo {author} {\bibfnamefont {T.~H.}\ \bibnamefont {Lo}}, \bibinfo {author} {\bibfnamefont {J.}~\bibnamefont {Qian}}, \bibinfo {author} {\bibfnamefont {A.}~\bibnamefont {Schleife}}, \bibinfo {author} {\bibfnamefont {J.~M.}\ \bibnamefont {Zuo}}, \bibinfo {author} {\bibfnamefont {V.}~\bibnamefont {Novosad}},\ and\ \bibinfo {author} {\bibfnamefont {A.}~\bibnamefont {Hoffmann}},\ }\bibfield  {title} {\bibinfo {title} {Integrating magnons for quantum information},\ }\href {https://doi.org/10.1063/5.0157520} {\bibfield  {journal} {\bibinfo  {journal} {Appl. Phys. Lett.}\ }\textbf {\bibinfo {volume} {123}},\ \bibinfo {eid} {130501} (\bibinfo {year} {2023})}\BibitemShut {NoStop}%
\bibitem [{\citenamefont {Baity}\ \emph {et~al.}(2021)\citenamefont {Baity}, \citenamefont {Bozhko}, \citenamefont {Macêdo}, \citenamefont {Smith}, \citenamefont {Holland}, \citenamefont {Danilin}, \citenamefont {Seferai}, \citenamefont {Barbosa}, \citenamefont {Peroor}, \citenamefont {Goldman}, \citenamefont {Nasti}, \citenamefont {Paul}, \citenamefont {Hadfield}, \citenamefont {McVitie},\ and\ \citenamefont {Weides}}]{Baity2021}%
  \BibitemOpen
  \bibfield  {author} {\bibinfo {author} {\bibfnamefont {P.~G.}\ \bibnamefont {Baity}}, \bibinfo {author} {\bibfnamefont {D.~A.}\ \bibnamefont {Bozhko}}, \bibinfo {author} {\bibfnamefont {R.}~\bibnamefont {Macêdo}}, \bibinfo {author} {\bibfnamefont {W.}~\bibnamefont {Smith}}, \bibinfo {author} {\bibfnamefont {R.~C.}\ \bibnamefont {Holland}}, \bibinfo {author} {\bibfnamefont {S.}~\bibnamefont {Danilin}}, \bibinfo {author} {\bibfnamefont {V.}~\bibnamefont {Seferai}}, \bibinfo {author} {\bibfnamefont {J.}~\bibnamefont {Barbosa}}, \bibinfo {author} {\bibfnamefont {R.~R.}\ \bibnamefont {Peroor}}, \bibinfo {author} {\bibfnamefont {S.}~\bibnamefont {Goldman}}, \bibinfo {author} {\bibfnamefont {U.}~\bibnamefont {Nasti}}, \bibinfo {author} {\bibfnamefont {J.}~\bibnamefont {Paul}}, \bibinfo {author} {\bibfnamefont {R.~H.}\ \bibnamefont {Hadfield}}, \bibinfo {author} {\bibfnamefont {S.}~\bibnamefont {McVitie}},\ and\ \bibinfo {author} {\bibfnamefont {M.}~\bibnamefont {Weides}},\ }\bibfield  {title} {\bibinfo {title}
  {Strong magnon-photon coupling with chip-integrated yig in the zero-temperature limit},\ }\href {https://doi.org/10.1063/5.0054837} {\bibfield  {journal} {\bibinfo  {journal} {Appl. Phys. Lett.}\ }\textbf {\bibinfo {volume} {119}},\ \bibinfo {eid} {033502} (\bibinfo {year} {2021})}\BibitemShut {NoStop}%
\bibitem [{\citenamefont {Li}\ \emph {et~al.}(2020)\citenamefont {Li}, \citenamefont {Zhang}, \citenamefont {Tyberkevych}, \citenamefont {Kwok}, \citenamefont {Hoffmann},\ and\ \citenamefont {Novosad}}]{Li2020}%
  \BibitemOpen
  \bibfield  {author} {\bibinfo {author} {\bibfnamefont {Y.}~\bibnamefont {Li}}, \bibinfo {author} {\bibfnamefont {W.}~\bibnamefont {Zhang}}, \bibinfo {author} {\bibfnamefont {V.}~\bibnamefont {Tyberkevych}}, \bibinfo {author} {\bibfnamefont {W.~K.}\ \bibnamefont {Kwok}}, \bibinfo {author} {\bibfnamefont {A.}~\bibnamefont {Hoffmann}},\ and\ \bibinfo {author} {\bibfnamefont {V.}~\bibnamefont {Novosad}},\ }\bibfield  {title} {\bibinfo {title} {Hybrid magnonics: Physics, circuits, and applications for coherent information processing},\ }\href {https://doi.org/10.1063/5.0020277} {\bibfield  {journal} {\bibinfo  {journal} {J. Appl. Phys.}\ }\textbf {\bibinfo {volume} {128}},\ \bibinfo {eid} {130902} (\bibinfo {year} {2020})}\BibitemShut {NoStop}%
\bibitem [{\citenamefont {Lachance-Quirion}\ \emph {et~al.}(2019)\citenamefont {Lachance-Quirion}, \citenamefont {Tabuchi}, \citenamefont {Gloppe}, \citenamefont {Usami},\ and\ \citenamefont {Nakamura}}]{Quirion2019}%
  \BibitemOpen
  \bibfield  {author} {\bibinfo {author} {\bibfnamefont {D.}~\bibnamefont {Lachance-Quirion}}, \bibinfo {author} {\bibfnamefont {Y.}~\bibnamefont {Tabuchi}}, \bibinfo {author} {\bibfnamefont {A.}~\bibnamefont {Gloppe}}, \bibinfo {author} {\bibfnamefont {K.}~\bibnamefont {Usami}},\ and\ \bibinfo {author} {\bibfnamefont {Y.}~\bibnamefont {Nakamura}},\ }\bibfield  {title} {\bibinfo {title} {Hybrid quantum systems based on magnonics},\ }\href {https://doi.org/10.7567/1882-0786/ab248d} {\bibfield  {journal} {\bibinfo  {journal} {Appl. Phys. Express}\ }\textbf {\bibinfo {volume} {12}},\ \bibinfo {eid} {070101} (\bibinfo {year} {2019})}\BibitemShut {NoStop}%
\bibitem [{\citenamefont {Forsch}\ \emph {et~al.}(2020)\citenamefont {Forsch}, \citenamefont {Stockill}, \citenamefont {Wallucks}, \citenamefont {Marinković}, \citenamefont {Gärtner}, \citenamefont {Norte}, \citenamefont {van Otten}, \citenamefont {Fiore}, \citenamefont {Srinivasan},\ and\ \citenamefont {Gröblacher}}]{Forsch2020}%
  \BibitemOpen
  \bibfield  {author} {\bibinfo {author} {\bibfnamefont {M.}~\bibnamefont {Forsch}}, \bibinfo {author} {\bibfnamefont {R.}~\bibnamefont {Stockill}}, \bibinfo {author} {\bibfnamefont {A.}~\bibnamefont {Wallucks}}, \bibinfo {author} {\bibfnamefont {I.}~\bibnamefont {Marinković}}, \bibinfo {author} {\bibfnamefont {C.}~\bibnamefont {Gärtner}}, \bibinfo {author} {\bibfnamefont {R.~A.}\ \bibnamefont {Norte}}, \bibinfo {author} {\bibfnamefont {F.}~\bibnamefont {van Otten}}, \bibinfo {author} {\bibfnamefont {A.}~\bibnamefont {Fiore}}, \bibinfo {author} {\bibfnamefont {K.}~\bibnamefont {Srinivasan}},\ and\ \bibinfo {author} {\bibfnamefont {S.}~\bibnamefont {Gröblacher}},\ }\bibfield  {title} {\bibinfo {title} {Microwave-to-optics conversion using a mechanical oscillator in its quantum ground state},\ }\href {https://doi.org/10.1038/s41567-019-0673-7} {\bibfield  {journal} {\bibinfo  {journal} {Nat. Phys.}\ }\textbf {\bibinfo {volume} {16}},\ \bibinfo {pages} {69} (\bibinfo {year} {2020})}\BibitemShut {NoStop}%
\bibitem [{\citenamefont {Kurizki}\ \emph {et~al.}(2015)\citenamefont {Kurizki}, \citenamefont {Bertet}, \citenamefont {Kubo}, \citenamefont {Mølmer}, \citenamefont {Petrosyan}, \citenamefont {Rabl},\ and\ \citenamefont {Schmiedmayer}}]{Kurizki2015}%
  \BibitemOpen
  \bibfield  {author} {\bibinfo {author} {\bibfnamefont {G.}~\bibnamefont {Kurizki}}, \bibinfo {author} {\bibfnamefont {P.}~\bibnamefont {Bertet}}, \bibinfo {author} {\bibfnamefont {Y.}~\bibnamefont {Kubo}}, \bibinfo {author} {\bibfnamefont {K.}~\bibnamefont {Mølmer}}, \bibinfo {author} {\bibfnamefont {D.}~\bibnamefont {Petrosyan}}, \bibinfo {author} {\bibfnamefont {P.}~\bibnamefont {Rabl}},\ and\ \bibinfo {author} {\bibfnamefont {J.}~\bibnamefont {Schmiedmayer}},\ }\bibfield  {title} {\bibinfo {title} {Quantum technologies with hybrid systems},\ }\href {https://doi.org/10.1073/pnas.1419326112} {\bibfield  {journal} {\bibinfo  {journal} {Proc. Natl. Acad. Sci. U. S. A.}\ }\textbf {\bibinfo {volume} {112}},\ \bibinfo {pages} {3866} (\bibinfo {year} {2015})}\BibitemShut {NoStop}%
\bibitem [{\citenamefont {Cao}\ \emph {et~al.}(2015)\citenamefont {Cao}, \citenamefont {Yan}, \citenamefont {Huebl}, \citenamefont {Goennenwein},\ and\ \citenamefont {Bauer}}]{Cao2015}%
  \BibitemOpen
  \bibfield  {author} {\bibinfo {author} {\bibfnamefont {Y.}~\bibnamefont {Cao}}, \bibinfo {author} {\bibfnamefont {P.}~\bibnamefont {Yan}}, \bibinfo {author} {\bibfnamefont {H.}~\bibnamefont {Huebl}}, \bibinfo {author} {\bibfnamefont {S.~T.}\ \bibnamefont {Goennenwein}},\ and\ \bibinfo {author} {\bibfnamefont {G.~E.}\ \bibnamefont {Bauer}},\ }\bibfield  {title} {\bibinfo {title} {Exchange magnon-polaritons in microwave cavities},\ }\href {https://doi.org/10.1103/PhysRevB.91.094423} {\bibfield  {journal} {\bibinfo  {journal} {Phys. Rev. B: Condens. Matter Mater. Phys.}\ }\textbf {\bibinfo {volume} {91}},\ \bibinfo {eid} {094423} (\bibinfo {year} {2015})}\BibitemShut {NoStop}%
\bibitem [{\citenamefont {Lachance-Quirion}\ \emph {et~al.}(2020)\citenamefont {Lachance-Quirion}, \citenamefont {Wolski}, \citenamefont {Tabuchi}, \citenamefont {Kono}, \citenamefont {Usami},\ and\ \citenamefont {Nakamura}}]{Nakamura2020}%
  \BibitemOpen
  \bibfield  {author} {\bibinfo {author} {\bibfnamefont {D.}~\bibnamefont {Lachance-Quirion}}, \bibinfo {author} {\bibfnamefont {S.~P.}\ \bibnamefont {Wolski}}, \bibinfo {author} {\bibfnamefont {Y.}~\bibnamefont {Tabuchi}}, \bibinfo {author} {\bibfnamefont {S.}~\bibnamefont {Kono}}, \bibinfo {author} {\bibfnamefont {K.}~\bibnamefont {Usami}},\ and\ \bibinfo {author} {\bibfnamefont {Y.}~\bibnamefont {Nakamura}},\ }\bibfield  {title} {\bibinfo {title} {Entanglement-based single-shot detection of a single magnon with a superconducting qubit},\ }\href {https://doi.org/10.1126/science.aaz9236} {\bibfield  {journal} {\bibinfo  {journal} {Science}\ }\textbf {\bibinfo {volume} {367}},\ \bibinfo {pages} {425} (\bibinfo {year} {2020})}\BibitemShut {NoStop}%
\bibitem [{\citenamefont {Tabuchi}\ \emph {et~al.}(2015)\citenamefont {Tabuchi}, \citenamefont {Ishino}, \citenamefont {Noguchi}, \citenamefont {Ishikawa}, \citenamefont {Yamazaki}, \citenamefont {Usami},\ and\ \citenamefont {Nakamura}}]{Tabuchi2015}%
  \BibitemOpen
  \bibfield  {author} {\bibinfo {author} {\bibfnamefont {Y.}~\bibnamefont {Tabuchi}}, \bibinfo {author} {\bibfnamefont {S.}~\bibnamefont {Ishino}}, \bibinfo {author} {\bibfnamefont {A.}~\bibnamefont {Noguchi}}, \bibinfo {author} {\bibfnamefont {T.}~\bibnamefont {Ishikawa}}, \bibinfo {author} {\bibfnamefont {R.}~\bibnamefont {Yamazaki}}, \bibinfo {author} {\bibfnamefont {K.}~\bibnamefont {Usami}},\ and\ \bibinfo {author} {\bibfnamefont {Y.}~\bibnamefont {Nakamura}},\ }\bibfield  {title} {\bibinfo {title} {Coherent coupling between a ferromagnetic magnon and a superconducting qubit},\ }\href {https://doi.org/10.1126/science.aaa3693} {\bibfield  {journal} {\bibinfo  {journal} {Science}\ }\textbf {\bibinfo {volume} {349}},\ \bibinfo {pages} {405} (\bibinfo {year} {2015})}\BibitemShut {NoStop}%
\bibitem [{\citenamefont {Knauer}\ \emph {et~al.}(2023)\citenamefont {Knauer}, \citenamefont {Davídková}, \citenamefont {Schmoll}, \citenamefont {Serha}, \citenamefont {Voronov}, \citenamefont {Wang}, \citenamefont {Verba}, \citenamefont {Dobrovolskiy}, \citenamefont {Lindner}, \citenamefont {Reimann}, \citenamefont {Dubs}, \citenamefont {Urbánek},\ and\ \citenamefont {Chumak}}]{Knauer2023}%
  \BibitemOpen
  \bibfield  {author} {\bibinfo {author} {\bibfnamefont {S.}~\bibnamefont {Knauer}}, \bibinfo {author} {\bibfnamefont {K.}~\bibnamefont {Davídková}}, \bibinfo {author} {\bibfnamefont {D.}~\bibnamefont {Schmoll}}, \bibinfo {author} {\bibfnamefont {R.~O.}\ \bibnamefont {Serha}}, \bibinfo {author} {\bibfnamefont {A.}~\bibnamefont {Voronov}}, \bibinfo {author} {\bibfnamefont {Q.}~\bibnamefont {Wang}}, \bibinfo {author} {\bibfnamefont {R.}~\bibnamefont {Verba}}, \bibinfo {author} {\bibfnamefont {O.~V.}\ \bibnamefont {Dobrovolskiy}}, \bibinfo {author} {\bibfnamefont {M.}~\bibnamefont {Lindner}}, \bibinfo {author} {\bibfnamefont {T.}~\bibnamefont {Reimann}}, \bibinfo {author} {\bibfnamefont {C.}~\bibnamefont {Dubs}}, \bibinfo {author} {\bibfnamefont {M.}~\bibnamefont {Urbánek}},\ and\ \bibinfo {author} {\bibfnamefont {A.~V.}\ \bibnamefont {Chumak}},\ }\bibfield  {title} {\bibinfo {title} {Propagating spin-wave spectroscopy in a liquid-phase epitaxial nanometer-thick yig film at millikelvin temperatures},\ }\href
  {https://doi.org/10.1063/5.0137437} {\bibfield  {journal} {\bibinfo  {journal} {J. Appl. Phys.}\ }\textbf {\bibinfo {volume} {133}},\ \bibinfo {eid} {143905} (\bibinfo {year} {2023})}\BibitemShut {NoStop}%
\bibitem [{\citenamefont {Kosen}\ \emph {et~al.}(2019)\citenamefont {Kosen}, \citenamefont {Loo}, \citenamefont {Bozhko}, \citenamefont {Mihalceanu},\ and\ \citenamefont {Karenowska}}]{Kosen2019}%
  \BibitemOpen
  \bibfield  {author} {\bibinfo {author} {\bibfnamefont {S.}~\bibnamefont {Kosen}}, \bibinfo {author} {\bibfnamefont {A.~F.~V.}\ \bibnamefont {Loo}}, \bibinfo {author} {\bibfnamefont {D.~A.}\ \bibnamefont {Bozhko}}, \bibinfo {author} {\bibfnamefont {L.}~\bibnamefont {Mihalceanu}},\ and\ \bibinfo {author} {\bibfnamefont {A.~D.}\ \bibnamefont {Karenowska}},\ }\bibfield  {title} {\bibinfo {title} {Microwave magnon damping in yig films at millikelvin temperatures},\ }\href {https://doi.org/10.1063/1.5115266} {\bibfield  {journal} {\bibinfo  {journal} {APL Mater.}\ }\textbf {\bibinfo {volume} {7}},\ \bibinfo {eid} {101120} (\bibinfo {year} {2019})}\BibitemShut {NoStop}%
\bibitem [{\citenamefont {Danilov}\ \emph {et~al.}(1989)\citenamefont {Danilov}, \citenamefont {Lyfar}, \citenamefont {Lyubonko}, \citenamefont {Nechiporuk},\ and\ \citenamefont {Ryabchenko}}]{Danilov1989}%
  \BibitemOpen
  \bibfield  {author} {\bibinfo {author} {\bibfnamefont {V.}~\bibnamefont {Danilov}}, \bibinfo {author} {\bibfnamefont {D.}~\bibnamefont {Lyfar}}, \bibinfo {author} {\bibfnamefont {Y.}~\bibnamefont {Lyubonko}}, \bibinfo {author} {\bibfnamefont {A.}~\bibnamefont {Nechiporuk}},\ and\ \bibinfo {author} {\bibfnamefont {S.}~\bibnamefont {Ryabchenko}},\ }\bibfield  {title} {\bibinfo {title} {Low-temperature ferromagnetic resonance in epitaxial garnet films on paramagnetic substrates},\ }\href {https://doi.org/10.1007/BF00897267} {\bibfield  {journal} {\bibinfo  {journal} {Russ. Phys. J.}\ }\textbf {\bibinfo {volume} {32}},\ \bibinfo {pages} {276} (\bibinfo {year} {1989})}\BibitemShut {NoStop}%
\bibitem [{\citenamefont {Guo}\ \emph {et~al.}(2022)\citenamefont {Guo}, \citenamefont {McCullian}, \citenamefont {Hammel},\ and\ \citenamefont {Yang}}]{Guo2022}%
  \BibitemOpen
  \bibfield  {author} {\bibinfo {author} {\bibfnamefont {S.}~\bibnamefont {Guo}}, \bibinfo {author} {\bibfnamefont {B.}~\bibnamefont {McCullian}}, \bibinfo {author} {\bibfnamefont {P.~C.}\ \bibnamefont {Hammel}},\ and\ \bibinfo {author} {\bibfnamefont {F.}~\bibnamefont {Yang}},\ }\bibfield  {title} {\bibinfo {title} {Low damping at few-{K} temperatures in {Y}$_{3}${Fe}$_{5}${O}$_{12}$ epitaxial films isolated from {Gd}$_{3}${Ga}$_{5}${O}$_{12}$ substrate using a diamagnetic {Y}$_{3}${Sc}$_{2.5}${Al}$_{2.5}${O}$_{12}$ spacer},\ }\href {https://doi.org/https://doi.org/10.1016/j.jmmm.2022.169795} {\bibfield  {journal} {\bibinfo  {journal} {J. Magn. Magn. Mater}\ }\textbf {\bibinfo {volume} {562}},\ \bibinfo {eid} {169795} (\bibinfo {year} {2022})}\BibitemShut {NoStop}%
\bibitem [{\citenamefont {Serha}\ \emph {et~al.}(2024)\citenamefont {Serha}, \citenamefont {Voronov}, \citenamefont {Schmoll}, \citenamefont {Verba}, \citenamefont {Levchenko}, \citenamefont {Koraltan}, \citenamefont {Davídková}, \citenamefont {Budinská}, \citenamefont {Wang}, \citenamefont {Dobrovolskiy}, \citenamefont {Urbánek}, \citenamefont {Lindner}, \citenamefont {Reimann}, \citenamefont {Dubs}, \citenamefont {Gonzalez-Ballestero}, \citenamefont {Abert}, \citenamefont {Suess}, \citenamefont {Bozhko}, \citenamefont {Knauer},\ and\ \citenamefont {Chumak}}]{Serha2024}%
  \BibitemOpen
  \bibfield  {author} {\bibinfo {author} {\bibfnamefont {R.~O.}\ \bibnamefont {Serha}}, \bibinfo {author} {\bibfnamefont {A.~A.}\ \bibnamefont {Voronov}}, \bibinfo {author} {\bibfnamefont {D.}~\bibnamefont {Schmoll}}, \bibinfo {author} {\bibfnamefont {R.}~\bibnamefont {Verba}}, \bibinfo {author} {\bibfnamefont {K.~O.}\ \bibnamefont {Levchenko}}, \bibinfo {author} {\bibfnamefont {S.}~\bibnamefont {Koraltan}}, \bibinfo {author} {\bibfnamefont {K.}~\bibnamefont {Davídková}}, \bibinfo {author} {\bibfnamefont {B.}~\bibnamefont {Budinská}}, \bibinfo {author} {\bibfnamefont {Q.}~\bibnamefont {Wang}}, \bibinfo {author} {\bibfnamefont {O.~V.}\ \bibnamefont {Dobrovolskiy}}, \bibinfo {author} {\bibfnamefont {M.}~\bibnamefont {Urbánek}}, \bibinfo {author} {\bibfnamefont {M.}~\bibnamefont {Lindner}}, \bibinfo {author} {\bibfnamefont {T.}~\bibnamefont {Reimann}}, \bibinfo {author} {\bibfnamefont {C.}~\bibnamefont {Dubs}}, \bibinfo {author} {\bibfnamefont {C.}~\bibnamefont {Gonzalez-Ballestero}}, \bibinfo {author}
  {\bibfnamefont {C.}~\bibnamefont {Abert}}, \bibinfo {author} {\bibfnamefont {D.}~\bibnamefont {Suess}}, \bibinfo {author} {\bibfnamefont {D.~A.}\ \bibnamefont {Bozhko}}, \bibinfo {author} {\bibfnamefont {S.}~\bibnamefont {Knauer}},\ and\ \bibinfo {author} {\bibfnamefont {A.~V.}\ \bibnamefont {Chumak}},\ }\bibfield  {title} {\bibinfo {title} {Magnetic anisotropy and ggg substrate stray field in yig films down to millikelvin temperatures},\ }\href {https://doi.org/10.1038/s44306-024-00030-7} {\bibfield  {journal} {\bibinfo  {journal} {npj Spintronics}\ }\textbf {\bibinfo {volume} {2}},\ \bibinfo {eid} {29} (\bibinfo {year} {2024})}\BibitemShut {NoStop}%
\bibitem [{\citenamefont {Boventer}\ \emph {et~al.}(2018)\citenamefont {Boventer}, \citenamefont {Pfirrmann}, \citenamefont {Krause}, \citenamefont {Sch\"on}, \citenamefont {Kl\"aui},\ and\ \citenamefont {Weides}}]{Weides2018}%
  \BibitemOpen
  \bibfield  {author} {\bibinfo {author} {\bibfnamefont {I.}~\bibnamefont {Boventer}}, \bibinfo {author} {\bibfnamefont {M.}~\bibnamefont {Pfirrmann}}, \bibinfo {author} {\bibfnamefont {J.}~\bibnamefont {Krause}}, \bibinfo {author} {\bibfnamefont {Y.}~\bibnamefont {Sch\"on}}, \bibinfo {author} {\bibfnamefont {M.}~\bibnamefont {Kl\"aui}},\ and\ \bibinfo {author} {\bibfnamefont {M.}~\bibnamefont {Weides}},\ }\bibfield  {title} {\bibinfo {title} {Complex temperature dependence of coupling and dissipation of cavity magnon polaritons from millikelvin to room temperature},\ }\href {https://doi.org/10.1103/PhysRevB.97.184420} {\bibfield  {journal} {\bibinfo  {journal} {Phys. Rev. B}\ }\textbf {\bibinfo {volume} {97}},\ \bibinfo {pages} {184420} (\bibinfo {year} {2018})}\BibitemShut {NoStop}%
\bibitem [{\citenamefont {Mihalceanu}\ \emph {et~al.}(2018)\citenamefont {Mihalceanu}, \citenamefont {Vasyuchka}, \citenamefont {Bozhko}, \citenamefont {Langner}, \citenamefont {Nechiporuk}, \citenamefont {Romanyuk}, \citenamefont {Hillebrands},\ and\ \citenamefont {Serga}}]{Mihalceanu2018}%
  \BibitemOpen
  \bibfield  {author} {\bibinfo {author} {\bibfnamefont {L.}~\bibnamefont {Mihalceanu}}, \bibinfo {author} {\bibfnamefont {V.~I.}\ \bibnamefont {Vasyuchka}}, \bibinfo {author} {\bibfnamefont {D.~A.}\ \bibnamefont {Bozhko}}, \bibinfo {author} {\bibfnamefont {T.}~\bibnamefont {Langner}}, \bibinfo {author} {\bibfnamefont {A.~Y.}\ \bibnamefont {Nechiporuk}}, \bibinfo {author} {\bibfnamefont {V.~F.}\ \bibnamefont {Romanyuk}}, \bibinfo {author} {\bibfnamefont {B.}~\bibnamefont {Hillebrands}},\ and\ \bibinfo {author} {\bibfnamefont {A.~A.}\ \bibnamefont {Serga}},\ }\bibfield  {title} {\bibinfo {title} {Temperature-dependent relaxation of dipole-exchange magnons in yttrium iron garnet films},\ }\href {https://doi.org/10.1103/PhysRevB.97.214405} {\bibfield  {journal} {\bibinfo  {journal} {Phys. Rev. B}\ }\textbf {\bibinfo {volume} {97}},\ \bibinfo {eid} {214405} (\bibinfo {year} {2018})}\BibitemShut {NoStop}%
\bibitem [{\citenamefont {Seiden}(1964)}]{Seiden1964}%
  \BibitemOpen
  \bibfield  {author} {\bibinfo {author} {\bibfnamefont {P.~E.}\ \bibnamefont {Seiden}},\ }\bibfield  {title} {\bibinfo {title} {Ferrimagnetic resonance relaxation in rare-earth iron garnets},\ }\href {https://doi.org/https://doi.org/10.1103/PhysRev.133.A728} {\bibfield  {journal} {\bibinfo  {journal} {Phys. Rev.}\ }\textbf {\bibinfo {volume} {133}},\ \bibinfo {eid} {A728} (\bibinfo {year} {1964})}\BibitemShut {NoStop}%
\bibitem [{\citenamefont {Sparks}\ \emph {et~al.}(1961)\citenamefont {Sparks}, \citenamefont {Loudon},\ and\ \citenamefont {Kittel}}]{Sparks1961}%
  \BibitemOpen
  \bibfield  {author} {\bibinfo {author} {\bibfnamefont {M.}~\bibnamefont {Sparks}}, \bibinfo {author} {\bibfnamefont {R.}~\bibnamefont {Loudon}},\ and\ \bibinfo {author} {\bibfnamefont {C.}~\bibnamefont {Kittel}},\ }\bibfield  {title} {\bibinfo {title} {Ferromagnetic relaxation. i. theory of the relaxation of the uniform precession and the degenerate spectrum in insulators at low temperatures},\ }\href {https://doi.org/10.1103/PhysRev.122.791} {\bibfield  {journal} {\bibinfo  {journal} {Phys. Rev.}\ }\textbf {\bibinfo {volume} {122}},\ \bibinfo {pages} {791} (\bibinfo {year} {1961})}\BibitemShut {NoStop}%
\bibitem [{\citenamefont {Dillon~Jr.}\ and\ \citenamefont {Nielsen}(1959)}]{Dillon1959}%
  \BibitemOpen
  \bibfield  {author} {\bibinfo {author} {\bibfnamefont {J.~F.}\ \bibnamefont {Dillon~Jr.}}\ and\ \bibinfo {author} {\bibfnamefont {J.~W.}\ \bibnamefont {Nielsen}},\ }\bibfield  {title} {\bibinfo {title} {Effects of rare earth impurities on ferrimagnetic resonance in yttrium iron garnet},\ }\href {https://doi.org/https://doi.org/10.1103/PhysRevLett.3.30} {\bibfield  {journal} {\bibinfo  {journal} {Phys. Rev. Lett.}\ }\textbf {\bibinfo {volume} {3}},\ \bibinfo {eid} {30} (\bibinfo {year} {1959})}\BibitemShut {NoStop}%
\bibitem [{\citenamefont {Spencer}\ \emph {et~al.}(1959)\citenamefont {Spencer}, \citenamefont {LeCraw},\ and\ \citenamefont {Clogston}}]{Spencer1959}%
  \BibitemOpen
  \bibfield  {author} {\bibinfo {author} {\bibfnamefont {E.~G.}\ \bibnamefont {Spencer}}, \bibinfo {author} {\bibfnamefont {R.~C.}\ \bibnamefont {LeCraw}},\ and\ \bibinfo {author} {\bibfnamefont {A.~M.}\ \bibnamefont {Clogston}},\ }\bibfield  {title} {\bibinfo {title} {Low-temperature line-width maximum in yttrium iron garnet},\ }\href {https://doi.org/https://doi.org/10.1103/PhysRevLett.3.32} {\bibfield  {journal} {\bibinfo  {journal} {Physical Review Letters}\ }\textbf {\bibinfo {volume} {3}},\ \bibinfo {eid} {32} (\bibinfo {year} {1959})}\BibitemShut {NoStop}%
\bibitem [{\citenamefont {Petrenko}\ \emph {et~al.}(1998)\citenamefont {Petrenko}, \citenamefont {Ritter}, \citenamefont {Yethiraj},\ and\ \citenamefont {McK~Paul}}]{Petrenko1998}%
  \BibitemOpen
  \bibfield  {author} {\bibinfo {author} {\bibfnamefont {O.~A.}\ \bibnamefont {Petrenko}}, \bibinfo {author} {\bibfnamefont {C.}~\bibnamefont {Ritter}}, \bibinfo {author} {\bibfnamefont {M.}~\bibnamefont {Yethiraj}},\ and\ \bibinfo {author} {\bibfnamefont {D.}~\bibnamefont {McK~Paul}},\ }\bibfield  {title} {\bibinfo {title} {Investigation of the low-temperature spin-liquid behavior of the frustrated magnet gadolinium gallium garnet},\ }\href {https://doi.org/10.1103/PhysRevLett.80.4570} {\bibfield  {journal} {\bibinfo  {journal} {Phys. Rev. Lett.}\ }\textbf {\bibinfo {volume} {80}},\ \bibinfo {pages} {4570} (\bibinfo {year} {1998})}\BibitemShut {NoStop}%
\bibitem [{\citenamefont {Tsui}\ \emph {et~al.}(1999)\citenamefont {Tsui}, \citenamefont {Kalechofsky}, \citenamefont {Burns},\ and\ \citenamefont {Schiffer}}]{Tsui1999}%
  \BibitemOpen
  \bibfield  {author} {\bibinfo {author} {\bibfnamefont {Y.~K.}\ \bibnamefont {Tsui}}, \bibinfo {author} {\bibfnamefont {N.}~\bibnamefont {Kalechofsky}}, \bibinfo {author} {\bibfnamefont {C.~A.}\ \bibnamefont {Burns}},\ and\ \bibinfo {author} {\bibfnamefont {P.}~\bibnamefont {Schiffer}},\ }\bibfield  {title} {\bibinfo {title} {Study of the low temperature thermal properties of the geometrically frustrated magnet: Gadolinium gallium garnet},\ }\href {https://doi.org/https://doi-org.uaccess.univie.ac.at/10.1063/1.370392} {\bibfield  {journal} {\bibinfo  {journal} {J. Appl. Phys.}\ }\textbf {\bibinfo {volume} {85}},\ \bibinfo {pages} {4512} (\bibinfo {year} {1999})}\BibitemShut {NoStop}%
\bibitem [{\citenamefont {Schiffer}\ \emph {et~al.}(1994)\citenamefont {Schiffer}, \citenamefont {Ramirez}, \citenamefont {Huse},\ and\ \citenamefont {Valentino}}]{Schiffer1994}%
  \BibitemOpen
  \bibfield  {author} {\bibinfo {author} {\bibfnamefont {P.}~\bibnamefont {Schiffer}}, \bibinfo {author} {\bibfnamefont {A.~P.}\ \bibnamefont {Ramirez}}, \bibinfo {author} {\bibfnamefont {D.~A.}\ \bibnamefont {Huse}},\ and\ \bibinfo {author} {\bibfnamefont {A.~J.}\ \bibnamefont {Valentino}},\ }\bibfield  {title} {\bibinfo {title} {Investigation of the field induced antiferromagnetic phase transition in the frustrated magnet: Gadolinium gallium garnet},\ }\href {https://doi.org/10.1103/PhysRevLett.73.2500} {\bibfield  {journal} {\bibinfo  {journal} {Phys. Rev. Lett.}\ }\textbf {\bibinfo {volume} {73}},\ \bibinfo {pages} {2500} (\bibinfo {year} {1994})}\BibitemShut {NoStop}%
\bibitem [{\citenamefont {Deen}\ \emph {et~al.}(2015)\citenamefont {Deen}, \citenamefont {Florea}, \citenamefont {Lhotel},\ and\ \citenamefont {Jacobsen}}]{Deen2015}%
  \BibitemOpen
  \bibfield  {author} {\bibinfo {author} {\bibfnamefont {P.~P.}\ \bibnamefont {Deen}}, \bibinfo {author} {\bibfnamefont {O.}~\bibnamefont {Florea}}, \bibinfo {author} {\bibfnamefont {E.}~\bibnamefont {Lhotel}},\ and\ \bibinfo {author} {\bibfnamefont {H.}~\bibnamefont {Jacobsen}},\ }\bibfield  {title} {\bibinfo {title} {Updating the phase diagram of the archetypal frustrated magnet ${\mathrm{gd}}_{3}{\mathrm{ga}}_{5}{\mathrm{o}}_{12}$},\ }\href {https://doi.org/10.1103/PhysRevB.91.014419} {\bibfield  {journal} {\bibinfo  {journal} {Phys. Rev. B}\ }\textbf {\bibinfo {volume} {91}},\ \bibinfo {pages} {014419} (\bibinfo {year} {2015})}\BibitemShut {NoStop}%
\bibitem [{\citenamefont {Cherepanov}\ \emph {et~al.}(1993)\citenamefont {Cherepanov}, \citenamefont {Kolokolov}, \citenamefont {L'vov},\ and\ \citenamefont {Cherepanop}}]{Cherepanov1993}%
  \BibitemOpen
  \bibfield  {author} {\bibinfo {author} {\bibfnamefont {V.}~\bibnamefont {Cherepanov}}, \bibinfo {author} {\bibfnamefont {I.}~\bibnamefont {Kolokolov}}, \bibinfo {author} {\bibfnamefont {V.}~\bibnamefont {L'vov}},\ and\ \bibinfo {author} {\bibfnamefont {V.}~\bibnamefont {Cherepanop}},\ }\bibfield  {title} {\bibinfo {title} {The saga of yig: Spectra, thermodynamics, interaction and relaxation of magnons in a complex magnet},\ }\href {https://doi.org/https://doi.org/10.1016/0370-1573(93)90107-O} {\bibfield  {journal} {\bibinfo  {journal} {Phys. Rep.}\ }\textbf {\bibinfo {volume} {229}},\ \bibinfo {pages} {81} (\bibinfo {year} {1993})}\BibitemShut {NoStop}%
\bibitem [{\citenamefont {Emtage}\ and\ \citenamefont {Daniel}(1984)}]{Emtage1984}%
  \BibitemOpen
  \bibfield  {author} {\bibinfo {author} {\bibfnamefont {P.~R.}\ \bibnamefont {Emtage}}\ and\ \bibinfo {author} {\bibfnamefont {M.~R.}\ \bibnamefont {Daniel}},\ }\bibfield  {title} {\bibinfo {title} {Magnetostatic waves and spin waves in layered ferrite structures},\ }\href {https://doi.org/10.1103/PhysRevB.29.212} {\bibfield  {journal} {\bibinfo  {journal} {Phys. Rev. B}\ }\textbf {\bibinfo {volume} {8}},\ \bibinfo {pages} {212} (\bibinfo {year} {1984})}\BibitemShut {NoStop}%
\bibitem [{\citenamefont {Barak}\ \emph {et~al.}(1992)\citenamefont {Barak}, \citenamefont {Huang},\ and\ \citenamefont {Bhagat}}]{Barak1992}%
  \BibitemOpen
  \bibfield  {author} {\bibinfo {author} {\bibfnamefont {J.}~\bibnamefont {Barak}}, \bibinfo {author} {\bibfnamefont {M.~X.}\ \bibnamefont {Huang}},\ and\ \bibinfo {author} {\bibfnamefont {S.~M.}\ \bibnamefont {Bhagat}},\ }\bibfield  {title} {\bibinfo {title} {Electron paramagnetic resonance study of gadolinium-gallium-garnet},\ }\href {https://doi.org/10.1063/1.351305} {\bibfield  {journal} {\bibinfo  {journal} {J. Appl. Phys.}\ }\textbf {\bibinfo {volume} {71}},\ \bibinfo {pages} {849} (\bibinfo {year} {1992})}\BibitemShut {NoStop}%
\bibitem [{\citenamefont {Bruckner}\ \emph {et~al.}(2012)\citenamefont {Bruckner}, \citenamefont {Vogler}, \citenamefont {Feischl}, \citenamefont {Praetorius}, \citenamefont {Bergmair}, \citenamefont {Huber}, \citenamefont {Fuger},\ and\ \citenamefont {Suess}}]{Dieter2012}%
  \BibitemOpen
  \bibfield  {author} {\bibinfo {author} {\bibfnamefont {F.}~\bibnamefont {Bruckner}}, \bibinfo {author} {\bibfnamefont {C.}~\bibnamefont {Vogler}}, \bibinfo {author} {\bibfnamefont {M.}~\bibnamefont {Feischl}}, \bibinfo {author} {\bibfnamefont {D.}~\bibnamefont {Praetorius}}, \bibinfo {author} {\bibfnamefont {B.}~\bibnamefont {Bergmair}}, \bibinfo {author} {\bibfnamefont {T.}~\bibnamefont {Huber}}, \bibinfo {author} {\bibfnamefont {M.}~\bibnamefont {Fuger}},\ and\ \bibinfo {author} {\bibfnamefont {D.}~\bibnamefont {Suess}},\ }\bibfield  {title} {\bibinfo {title} {3d fem–bem-coupling method to solve magnetostatic maxwell equations},\ }\href {https://doi.org/https://doi.org/10.1016/j.jmmm.2012.01.016} {\bibfield  {journal} {\bibinfo  {journal} {J. Magn. Magn. Mater.}\ }\textbf {\bibinfo {volume} {324}},\ \bibinfo {pages} {1862} (\bibinfo {year} {2012})}\BibitemShut {NoStop}%
\bibitem [{\citenamefont {Bruckner}\ \emph {et~al.}(2023)\citenamefont {Bruckner}, \citenamefont {Koraltan}, \citenamefont {Abert},\ and\ \citenamefont {Suess}}]{bruckner2023magnum}%
  \BibitemOpen
  \bibfield  {author} {\bibinfo {author} {\bibfnamefont {F.}~\bibnamefont {Bruckner}}, \bibinfo {author} {\bibfnamefont {S.}~\bibnamefont {Koraltan}}, \bibinfo {author} {\bibfnamefont {C.}~\bibnamefont {Abert}},\ and\ \bibinfo {author} {\bibfnamefont {D.}~\bibnamefont {Suess}},\ }\bibfield  {title} {\bibinfo {title} {Magnum.np: a pytorch based gpu enhanced finite difference micromagnetic simulation framework for high level development and inverse design},\ }\href {https://doi.org/https://doi.org/10.1038/s41598-023-39192-5} {\bibfield  {journal} {\bibinfo  {journal} {Scientific Reports}\ }\textbf {\bibinfo {volume} {13}},\ \bibinfo {eid} {12054} (\bibinfo {year} {2023})}\BibitemShut {NoStop}%
\bibitem [{\citenamefont {Kalarickal}\ \emph {et~al.}(2006)\citenamefont {Kalarickal}, \citenamefont {Krivosik}, \citenamefont {Wu}, \citenamefont {Patton}, \citenamefont {Schneider}, \citenamefont {Kabos}, \citenamefont {Silva},\ and\ \citenamefont {Nibarger}}]{kalarickal2006ferromagnetic}%
  \BibitemOpen
  \bibfield  {author} {\bibinfo {author} {\bibfnamefont {S.~S.}\ \bibnamefont {Kalarickal}}, \bibinfo {author} {\bibfnamefont {P.}~\bibnamefont {Krivosik}}, \bibinfo {author} {\bibfnamefont {M.}~\bibnamefont {Wu}}, \bibinfo {author} {\bibfnamefont {C.~E.}\ \bibnamefont {Patton}}, \bibinfo {author} {\bibfnamefont {M.~L.}\ \bibnamefont {Schneider}}, \bibinfo {author} {\bibfnamefont {P.}~\bibnamefont {Kabos}}, \bibinfo {author} {\bibfnamefont {T.~J.}\ \bibnamefont {Silva}},\ and\ \bibinfo {author} {\bibfnamefont {J.~P.}\ \bibnamefont {Nibarger}},\ }\bibfield  {title} {\bibinfo {title} {Ferromagnetic resonance linewidth in metallic thin films: Comparison of measurement methods},\ }\href {https://doi.org/https://doi.org/10.1063/1.2197087} {\bibfield  {journal} {\bibinfo  {journal} {J. Appl. Phys.}\ }\textbf {\bibinfo {volume} {99}},\ \bibinfo {eid} {093909} (\bibinfo {year} {2006})}\BibitemShut {NoStop}%
\bibitem [{\citenamefont {Kasuya}\ and\ \citenamefont {LeCraw}(1961)}]{Kasuya1961}%
  \BibitemOpen
  \bibfield  {author} {\bibinfo {author} {\bibfnamefont {T.}~\bibnamefont {Kasuya}}\ and\ \bibinfo {author} {\bibfnamefont {R.~C.}\ \bibnamefont {LeCraw}},\ }\bibfield  {title} {\bibinfo {title} {Relaxation mechanisms in ferromagnetic resonance},\ }\href {https://doi.org/https://doi.org/10.1103/PhysRevLett.6.223} {\bibfield  {journal} {\bibinfo  {journal} {Phys. Rev. Lett.}\ }\textbf {\bibinfo {volume} {6}},\ \bibinfo {eid} {223} (\bibinfo {year} {1961})}\BibitemShut {NoStop}%
\bibitem [{\citenamefont {Tabuchi}\ \emph {et~al.}(2014)\citenamefont {Tabuchi}, \citenamefont {Ishino}, \citenamefont {Ishikawa}, \citenamefont {Yamazaki}, \citenamefont {Usami},\ and\ \citenamefont {Nakamura}}]{Tabuchi2014}%
  \BibitemOpen
  \bibfield  {author} {\bibinfo {author} {\bibfnamefont {Y.}~\bibnamefont {Tabuchi}}, \bibinfo {author} {\bibfnamefont {S.}~\bibnamefont {Ishino}}, \bibinfo {author} {\bibfnamefont {T.}~\bibnamefont {Ishikawa}}, \bibinfo {author} {\bibfnamefont {R.}~\bibnamefont {Yamazaki}}, \bibinfo {author} {\bibfnamefont {K.}~\bibnamefont {Usami}},\ and\ \bibinfo {author} {\bibfnamefont {Y.}~\bibnamefont {Nakamura}},\ }\bibfield  {title} {\bibinfo {title} {Hybridizing ferromagnetic magnons and microwave photons in the quantum limit},\ }\href {https://doi.org/10.1103/PhysRevLett.113.083603} {\bibfield  {journal} {\bibinfo  {journal} {Phys. Rev. Lett.}\ }\textbf {\bibinfo {volume} {113}},\ \bibinfo {pages} {083603} (\bibinfo {year} {2014})}\BibitemShut {NoStop}%
\bibitem [{\citenamefont {Matula}(1979)}]{Matula1979}%
  \BibitemOpen
  \bibfield  {author} {\bibinfo {author} {\bibfnamefont {R.~A.}\ \bibnamefont {Matula}},\ }\bibfield  {title} {\bibinfo {title} {Electrical resistivity of copper, gold, palladium, and silver},\ }\href {https://doi.org/10.1063/1.555614} {\bibfield  {journal} {\bibinfo  {journal} {J. Phys. Chem. Ref. Data}\ }\textbf {\bibinfo {volume} {8}},\ \bibinfo {pages} {1147} (\bibinfo {year} {1979})}\BibitemShut {NoStop}%
\bibitem [{\citenamefont {Corporation}(2019)}]{datasheetDuroid}%
  \BibitemOpen
  \bibfield  {author} {\bibinfo {author} {\bibfnamefont {R.}~\bibnamefont {Corporation}},\ }\href@noop {} {\bibinfo {title} {Rt/duroid 6010.2lm laminates datasheet}} (\bibinfo {year} {2019}),\ \bibinfo {note} {available: \url{https://rogerscorp.com/documents/advanced-electronics-solutions/english/data-sheets/rt-duroid-6006-6010lm-laminate-data-sheet.pdf}}\BibitemShut {NoStop}%
\bibitem [{\citenamefont {Kalinikos}\ and\ \citenamefont {Slavin}(1986)}]{Kalinikos1986}%
  \BibitemOpen
  \bibfield  {author} {\bibinfo {author} {\bibfnamefont {B.}~\bibnamefont {Kalinikos}}\ and\ \bibinfo {author} {\bibfnamefont {A.}~\bibnamefont {Slavin}},\ }\bibfield  {title} {\bibinfo {title} {Theory of dipole-exchange spin wave spectrum for ferromagnetic films with mixed exchange boundary conditions},\ }\href {https://doi.org/10.1088/0022-3719/19/35/014} {\bibfield  {journal} {\bibinfo  {journal} {J. Phys. C: Solid State Phys.}\ }\textbf {\bibinfo {volume} {19}},\ \bibinfo {pages} {7013} (\bibinfo {year} {1986})}\BibitemShut {NoStop}%
\end{thebibliography}%

\end{document}